\documentclass[a4paper,usenames,dvipsnames,11pt]{article}
\pdfoutput=1

\usepackage{jheppub}
\usepackage{slashed}
\usepackage{mathrsfs,booktabs,multirow,tabularx}
\usepackage{stmaryrd}
\usepackage{xspace}
\usepackage{fancyvrb}
\usepackage[makeroom]{cancel}
\usepackage{amsmath}    
\usepackage{amssymb}    %
\usepackage{graphicx}   
\usepackage{verbatim}   
\usepackage{lscape}
\usepackage{subfig}
\usepackage{listings}
\usepackage{mathtools}
\usepackage{enumerate}

\def\beq{\begin{equation}}
\def\eeq{\end{equation}}
\def\beqn{\begin{eqnarray}}
\def\eeqn{\end{eqnarray}}

\newcommand{\bqa}{\begin{eqnarray}}
\newcommand{\eqa}{\end{eqnarray}}
\chardef\MyArticleWithColor=\pdfcolorstackinit page direct{0 g}

\def\cCode#1{\begin{lstlisting}[mathescape,basicstyle=\small
\ttfamily,frame=leftline,aboveskip=4mm,belowskip=4mm,xleftmargin=20pt,framexleftmargin=10pt,
numbers=none,framerule=2pt,abovecaptionskip=0.0mm,belowcaptionskip=3.5mm #1]}

\newcommand\sss{\scriptscriptstyle}
\newcommand\alphas{\alpha_s}
\newcommand\yt{y_t}
\newcommand\tth{t\bar t H}

\newcommand{\tev}{\,\textrm{TeV}}
\newcommand{\gev}{\,\textrm{GeV}}
\newcommand{\fb}{\textrm{fb}}

\newcommand\muF{\mu_{\sss F}}
\newcommand\muR{\mu_{\sss R}}
\newcommand\muIR{\mu_{\sss IR}}

\newcommand{\mt}{m_{t}}

\def\beq{\begin{equation}}
\def\eeq{\end{equation}}
\def\beqar{\begin{eqnarray}}
\def\eeqar{\end{eqnarray}}
\def\barr#1{\begin{array}{#1}}
\def\earr{\end{array}}
\def\bfi{\begin{figure}}
\def\efi{\end{figure}}
\def\btab{\begin{table}}
\def\etab{\end{table}}
\def\bce{\begin{center}}
\def\ece{\end{center}}
\def\nn{\nonumber}

\def\si{\sigma}

\newcommand{\TeV}{\unskip\,\mathrm{TeV}}
\newcommand{\GeV}{\unskip\,\mathrm{GeV}}

\newcommand{\M}{{\cal{M}}}

\def\mathswitch#1{\relax\ifmmode#1\else$#1$\fi}

\usepackage{comment}

\title{State-of-the-art cross sections for $\boldsymbol{t\bar t H}$:\\ NNLO predictions matched with NNLL resummation and EW corrections}

\author[1]{Roger Balsach}
\author[2]{Alessandro Broggio}
\author[3]{Simone Devoto}
\author[4]{Andrea Ferroglia}
\author[5]{Rikkert Frederix}
\author[6]{Massimiliano Grazzini}
\author[6]{Stefan Kallweit}
\author[1]{Anna Kulesza}
\author[7]{Javier Mazzitelli}
\author[8]{Leszek Motyka}
\author[9]{Davide Pagani}
\author[10]{Benjamin D. Pecjak}
\author[11]{Chiara Savoini}
\author[8]{Tomasz Stebel}
\author[12]{Malgorzata Worek}
\author[13]{Marco Zaro}

\affiliation[1]{Institute for Theoretical Physics, University of M\"unster, D-48149 M\"unster, Germany}
\affiliation[2]{Faculty of Physics, University of Vienna, Boltzmanngasse 5, A-1090 Vienna, Austria}
\affiliation[3]{Department of Physics and Astronomy, Ghent University, 9000 Ghent, Belgium}
\affiliation[4]{Physics Department, New York City College of Technology, 
The City University of New York, 300 Jay Street, Brooklyn, NY 11201, USA \&
The Graduate School and University Center, The City University of New York, 365 Fifth Avenue, New York, NY 10016, USA
}
\affiliation[5]{Department of Physics, Lund University, SE-223 63 Lund, Sweden}
\affiliation[6]{Physik Institut, Universit\"at Z\"urich, 8057 Z\"urich, Switzerland}
\affiliation[7]{Paul Scherrer Institut, 5232 Villigen PSI, Switzerland}
\affiliation[8]{Institute of Theoretical Physics, Jagellonian University, S.\L{}ojasiewicza 11, 30-348 Krak\'ow, Poland}
\affiliation[9]{INFN, Sezione di Bologna, Via Irnerio 46, 40126 Bologna, Italy}
\affiliation[10]{Institute for Particle Physics Phenomenology, Department of Physics Durham University, Durham DH1 3LE, United Kingdom}
\affiliation[11]{Technical University of Munich, TUM School of Natural Sciences, Physics Department, James-Franck-Stra$\ss$e 1, 85748 Garching, Germany}
\affiliation[12]{Institute for Theoretical Particle Physics and Cosmology, RWTH Aachen University, D-52056 Aachen, Germany}
\affiliation[13]{TIFLab, Universit\`a degli Studi di Milano \& INFN, Sezione di Milano, Via Celoria 16, 20133 Milano, Italy}

\emailAdd{rbalsach@uni-muenster.de}
\emailAdd{alessandro.broggio@univie.ac.at}
\emailAdd{simone.devoto@ugent.be}
\emailAdd{aferroglia@citytech.cuny.edu}
\emailAdd{rikkert.frederix@fysik.lu.se}
\emailAdd{grazzini@physik.uzh.ch}
\emailAdd{stefan.kallweit@physik.uzh.ch}
\emailAdd{anna.kulesza@uni-muenster.de}
\emailAdd{javier.mazzitelli@psi.ch}
\emailAdd{leszekm@th.if.uj.edu.pl}
\emailAdd{davide.pagani@bo.infn.it}
\emailAdd{chiara.savoini@tum.de}
\emailAdd{tomasz.stebel@uj.edu.pl}
\emailAdd{worek@physik.rwth-aachen.de}
\emailAdd{marco.zaro@mi.infn.it}

\abstract{We present new, state-of-the-art predictions for the associated production
of the SM Higgs boson with top quarks, computed in accordance with the
recommendations of the LHC Higgs Working Group.
The NNLO QCD predictions, derived through suitable approximations of the two-loop virtual contribution, are supplemented with soft-gluon resummation up to NNLL accuracy.
Two distinct resummation frameworks are employed -- one based on direct QCD and the
other on soft-collinear effective theory -- and their features are
compared in detail. These results are further combined with the complete-NLO corrections, yielding the most precise SM predictions for this process to date. The relevant sources of theoretical uncertainties are thoroughly estimated and discussed.}
\preprint{
\begin{flushright}

LHCHWG-2025-001,\\
IPPP/25/01, TUM-HEP-1549/25, UWThPh 2024-25, \\
TTK-25-01, P3H-25-001, TIF-UNIMI-2025-3,\\
MS-TP-25-02, PSI-PR-25-03, ZU-TH 08/25

\end{flushright}
}

\begin{document}
\maketitle
\flushbottom

%
\section{Introduction}
\label{sec:intro}
%
%

After the discovery of the Higgs boson at the Large Hadron Collider
(LHC) at CERN~\cite{Aad:2012tfa,Chatrchyan:2012ufa}, an unprecedented
campaign of measurements has begun. Such a campaign aims to
thoroughly scrutinise the properties of the newly discovered particle,
on the one hand to verify the consistency of the Standard Model (SM)
of fundamental interactions, and on the other to exploit this particle
as a magnifying glass for new, yet undiscovered physics phenomena.
More than a decade after its discovery and on the cusp of the LHC Run
III, the SM description of the Higgs sector has been corroborated at
the level of a few percent. Indeed, the new combination of all the Run
II data yields $\mu = 1.002 \pm 0.036 \, ({\rm th.}) \pm 0.029 \,
({\rm stat.}) \pm 0.033 \, ({\rm syst.}) = 1.002\pm 0.057$
\cite{ATLAS:2022vkf,CMS:2022dwd}, where $\mu$ is a common
signal-strength parameter that quantifies the agreement between the
observed signal yields from all production modes and decay channels,
and the corresponding SM expectations.  The uncertainties associated
with the new measurement reflect a $4.5$-fold improvement in precision
compared to that achieved at the time of discovery. Currently, the
theoretical uncertainties in both signal and background modelling as
well as the experimental statistical and systematic uncertainties
contribute at a similar level. Achieving this level of accuracy, as well as matching 
that of the Run III data, requires continuously improving theoretical predictions 
for the various Higgs boson production processes and their backgrounds.

Among the studied properties of the Higgs boson, its coupling to
fermions (quark and leptons) has been among the most recently
established.  Experimental evidence for the Higgs boson interaction
with third-generation fermions (top and bottom quarks, and the $\tau$
lepton) was obtained only six years after the Higgs discovery
\cite{CMS:2014wdm,CMS:2017odg,ATLAS:2017ztq,CMS:2018fdh,CMS:2018uxb,
ATLAS:2018mme,CMS:2018nsn,ATLAS:2018kot,ATLAS:2018ynr}, while evidence
for its coupling to second-generation fermions followed in 2020, with
the observation of the Higgs boson interaction with
muons~\cite{CMS:2020xwi}. More recently, tighter constraints have been
placed on the charm-quark Yukawa coupling by the ATLAS and CMS
collaborations \cite{CMS:2022psv,ATLAS:2024yzu}. The importance of
probing the Higgs couplings to fermions stems from the fact that such
couplings are mediated by the so-called Yukawa interaction, a novel
type of interaction that has never been observed among elementary
particles before. Furthermore, since the known fermions acquire their
masses via the same interaction, the Higgs-fermion interactions can
shed some light on the observed pattern of quark and lepton masses in
the SM.


Of all fermions, the top quark is the heaviest and, consequently, the
most strongly coupled to the Higgs boson. The top-quark Yukawa
coupling, $\yt$, can be probed indirectly via a number of processes: 
the Higgs gluon-fusion, where the top quark appears in a closed
loop, the four top production  $pp\to t\bar{t}t\bar{t}$,
which  is sensitive to $\yt$ at the tree level via diagrams featuring
an off-shell Higgs propagator \cite{Cao:2016wib,Cao:2019ygh,
  CMS:2019rvj}, and also the top-pair production $pp\to t\bar{t}$,
where high-precision measurements allow sensitivity to
effects induced by $\yt$ via EW loops \cite{Kuhn:2013zoa,
  Martini:2019lsi, CMS:2020djy, Martini:2021uey, Maltoni:2024wyh}.  In
all these cases, the extraction of $\yt$ depends
on the assumptions made about new-physics effects, particularly on the (non-)existence of
new particles which couple to the Higgs boson or to the top
quark. Still, such processes can be exploited to probe the ${\cal CP}$
properties of the top-Higgs interaction, similarly to the case of
single-top Higgs production \cite{Demartin:2015uha, Demartin:2016axk},
which unlike the aforementioned processes features  a Higgs boson in
the final state.

On the other hand, a more direct and
model-independent measurement of $\yt$ can instead be achieved by
analysing the $ pp\to t\bar{t}H$ production process. This approach was
instrumental in providing the first evidence of the Higgs-top
coupling, as reported in Refs.~\cite{ATLAS:2017ztq,CMS:2018fdh}. Besides the magnitude of $y_t$, also
the ${\cal{CP}}$ structure of the top-Higgs interaction has been studied at the LHC
with astonishing scrutiny. Both ATLAS and CMS have reported Higgs-top
${\cal CP}$ studies, investigating the $pp\to t{t}H$ process with
different Higgs boson decay channels. Although the LHC measurements
support the SM $\yt$ coupling, a ${\cal CP}$-violating coupling has
not yet been ruled out \cite{ATLAS:2020ior,CMS:2020cga,CMS:2022dbt,
ATLAS:2022tnm,ATLAS:2023cbt,CMS:2024fdo}. The presence of the latter
would of course mean a departure from the SM predictions and evidence
for new physics effects.

The computation of theoretical predictions including higher-order
effects in perturbation theory for the $pp \to t\bar{t}H$ process with
stable top quarks began two decades ago. The first computations at
next-to-leading order (NLO) in QCD have been carried out by two
independent groups
\cite{Beenakker:2001rj,Reina:2001sf,Reina:2001bc,Beenakker:2002nc,
Dawson:2002tg}. About one decade later, the NLO EW
corrections~\cite{Frixione:2014qaa,Zhang:2014gcy,Frixione:2015zaa},
and the so-called complete-NLO predictions have been computed
\cite{Frederix:2018nkq}. These complete theoretical predictions
comprise all leading and subleading LO contributions as well as their
corresponding higher-order QCD and EW effects. The complexity of the
final state, with three massive particles, has prevented the
computation of next-to-next-to-leading order (NNLO) predictions for a
long time.  A first step was made in Ref.~\cite{Catani:2021cbl} where
NNLO QCD corrections for the flavour off-diagonal partonic channels
were computed. Only recently, a complete NNLO calculation, including
the diagonal partonic channels, has been presented in
Ref.~\cite{Catani:2022mfv}, albeit with a soft Higgs boson
approximation for the two-loop amplitudes.  This approach has been
further improved in Ref.~\cite{Devoto:2024nhl} where the soft Higgs
boson approximation is combined with a high-energy expansion in the
small top-mass limit.  Despite substantial recent advancements, the
quest for the calculation of the full two-loop amplitude for the $t
\bar t H$ process is still ongoing, see
e.g. Refs.~\cite{FebresCordero:2023pww,Agarwal:2024jyq,Wang:2024pmv}.
In Ref.~\cite{Devoto:2024nhl}, the newly computed NNLO QCD results,
based on the approximated double-virtual contribution, have also been
equipped with the complete set of EW corrections.  Before NNLO
predictions became available, higher-order effects have been estimated
via resummation techniques, notably for those contributions arising
from soft-gluon emissions. These have been computed by different
groups up to next-to-next-to-leading logarithmic (NNLL) accuracy
\cite{Kulesza:2015vda,Broggio:2015lya,Broggio:2016lfj,
Kulesza:2017ukk,Ju:2019lwp}, and have subsequently been matched to the
complete-NLO predictions
\cite{Kulesza:2018tqz,Broggio:2019ewu,Kulesza:2020nfh}.  The $pp\to
t\bar{t}H$ production process has also been examined in the Standard
Model Effective Field Theory at the NLO level in QCD, see e.g.\
Refs. \cite{Maltoni:2016yxb,DiNoi:2023onw}.

For what concerns the simulation of various differential cross-section
distributions, NLO QCD predictions for the $pp \to t\bar{t}H$ process
matched to parton-shower simulations have been available for over a
decade \cite{Frederix:2011zi,Hartanto:2015uka}. More recently, the
simulation of full off-shell effects at NLO QCD
\cite{Denner:2015yca,Stremmer:2021bnk} and also at NLO EW
\cite{Denner:2016wet} have been attained. All
resonant and non-resonant Feynman diagrams, interferences, and
finite-width effects of the top quarks and $W^\pm/Z$ gauge bosons have
been included in these calculations. In practice, higher-order QCD and EW corrections have
been calculated for the final state $\ell^+\nu_\ell \, \ell^-
\bar{\nu}_\ell \, b\bar{b} \,H$, where $\ell^\pm=e^\pm,\mu^\pm$. The
results presented in Ref.~\cite{Stremmer:2021bnk} also incorporated
effects due to decays of the Higgs boson, albeit in the narrow-width
approximation.  In addition, NLO QCD predictions are available in the
literature, based on either on-shell or full off-shell modelling of
top-quark decays, which take into account the mixing between the Higgs
boson's ${\cal CP}$-even and ${\cal CP}$-odd states for various
observables
\cite{Demartin:2014fia,Demartin:2015uha,Demartin:2016axk,Hermann:2022vit}.

In this work, we provide cross-section predictions for the $pp \to
t\bar{t}H$ process which represent the new state-of-the-art for what
concerns the inclusion of higher-order effects.  Our computation
builds upon the NNLO QCD result recently presented in
Ref.~\cite{Devoto:2024nhl} and derived through suitable approximations
of the genuine two-loop contribution.  This result is supplemented by
soft-gluon resummation up to NNLL accuracy and by the complete-NLO 
corrections. Besides the theoretical prediction, we provide a
comprehensive estimate of the theoretical uncertainties due to missing
higher-order effects, parton distribution functions, the strong
coupling $\alphas$ and the top-quark mass $\mt$.

The rest of the paper is organised as follows.  In
section \ref{sec:theo}, we review the theoretical framework and summarise
the various contributions that enter our state-of-the-art
predictions. In section \ref{sec:results} we present our results. Our
conclusions are given in section \ref{sec:concl}. Finally, in section
\ref{sec:cite} we present our citation policy, which should be
followed when the results of our work are used in other scientific
papers.

%
\section{Theoretical predictions for $t\bar t H$}
\label{sec:theo}
%

\def\t{{\bar t}}
\def\Mtt{m(t\bar t)}
\def\PTtt{p_{T,t\bar t}}
\def\PTt{p_{T,t}}
\def\PTtbar{p_{T,{\bar t}}}
\def\PTavt{p_{T,{\rm avt}}}
\def\Yavt{y_{\rm avt}}
\def\Ytt{y(t\bar t)}
\def\GeV{\, \rm GeV}
\def\TeV{\, \rm TeV}
\def\eps{\varepsilon}
\def\alphas{\alpha_s}
\def\SigmaSub{\Sigma_{\rm sub}}

In this section, we discuss the technical details for the various
contributions entering our predictions at NNLO+NNLL accuracy,
including the complete-NLO corrections.  We will also introduce the
naming conventions used throughout the paper. Starting with QCD
effects, in section \ref{sec:theo-nnlo} we report on the computation
of the NNLO corrections, while in section \ref{sec:theo-nnllb}  and
section \ref{sec:theo-nnllk} we present the two resummation frameworks
employed, namely soft-collinear effective theory (SCET) and direct QCD
(dQCD). These two methods are compared in section
\ref{sec:theo-nnllcomp}, while the inclusion of EW effects is
discussed in section \ref{sec:theo-ew}. The reader not interested in the
technical details may skip directly to section \ref{sec:names}, where we
outline the naming conventions for predictions computed at different
accuracies and explain how they are combined to obtain our
state-of-the-art predictions.

\subsection{NNLO predictions}
\label{sec:theo-nnlo}
It is well known that a bottleneck in performing NNLO QCD calculations is the availability of the corresponding two-loop amplitudes.
This is particularly true for processes beyond the $2\to2$ scattering topology, especially when several mass scales are involved.
Despite the significant progress achieved in the last few years in multiloop computations (see e.g. Ref.~\cite{Andersen:2024czj} and references therein), to date, exact two-loop
amplitudes for $t\bar{t}H$ production are still unavailable.
For practical phenomenological applications,
a promising strategy consists in obtaining the double-virtual contribution in some approximate form while keeping the rest of the NNLO calculation exact.

Besides the treatment of the two-loop amplitudes, the mediation of infrared~(IR) singularities between the different amplitudes and phase spaces
is also a challenging task. In the NNLO calculation of Refs.~\cite{Catani:2022mfv,Devoto:2024nhl} the transverse-momentum~($q_T$) subtraction method~\cite{Catani:2007vq} was employed.
This method uses IR subtraction counterterms that are constructed by considering
the $q_T$ distribution
of the produced final-state system in the limit \mbox{$q_T \to 0$}~\cite{Catani:2013tia,Zhu:2012ts,Li:2013mia,Catani:2014qha}.
Originally developed for the production of a colour singlet system, the method has been extended to heavy-quark production~\cite{Bonciani:2015sha,Catani:2023tby} and
applied to the NNLO computations of top-quark and bottom-quark pair production~\cite{Catani:2019iny,Catani:2019hip,Catani:2020kkl}.

The production of a heavy-quark pair accompanied by a colourless particle does not pose any additional conceptual complications in the context of the $q_T$ subtraction formalism.
However, its implementation requires the computation of appropriate soft-parton contributions.
The results of this computation at NLO and, partly, at NNLO were presented in Ref.~\cite{Catani:2021cbl},
and the evaluation of the NNLO soft terms has been subsequently completed~\cite{inprepQQXsoft} and applied
to several associated heavy-quark pair production processes~\cite{Catani:2022mfv,Buonocore:2022pqq,Buonocore:2023ljm,Devoto:2024nhl}.

Following the NNLO computation of the off-diagonal partonic channels~\cite{Catani:2021cbl},
a first complete NNLO result for $t {\bar t}H$ production was presented in Ref.~\cite{Catani:2022mfv}, where a purely soft Higgs boson approximation
was developed to estimate the yet unknown two-loop amplitudes.
A step forward was recently made in Ref.~\cite{Devoto:2024nhl}, where the first fully differential results for $t {\bar t}H$ production were presented and a complementary and rather different approximation of the two-loop amplitudes was introduced. This approximation is based on a high-energy (or small top-quark mass) expansion. 

To isolate the part of the NNLO calculation that requires an approximation, 
the two-loop hard-virtual coefficient is defined as
\begin{equation}
\label{eq:H}
	H^{(2)}(\muIR)=\left.\frac{2{\mathrm{Re}}\left(\M^{(2), \mathrm{fin}}(\muIR,\muR)\M^{(0)*}\right)}{|\M^{(0)}|^2}\right\vert_{\muR=Q}\,,
\end{equation}
and is computed through the interference of the Born amplitude ${\cal M}^{(0)}$ for the \mbox{$c{\bar c}\to t{\bar t}H$} process (\mbox{$c=q,g$})
with the IR-subtracted two-loop amplitude \mbox{$\M^{(2), \mathrm{fin}}(\muIR,\muR)$} (in an expansion in powers of $\alphas/(2 \pi)$).
To be precise, \mbox{$\M^{(2), \mathrm{fin}}(\muIR,\muR)$} is evaluated within the scheme of Ref.~\cite{Ferroglia:2009ii} at the subtraction scale $\muIR$.
The central value of $\muIR$ is set to the invariant mass $Q$ of the event.
The ensuing contribution of the $H^{(2)}$ coefficient to the NNLO cross section reads
\begin{equation}
\label{eq:sigma2}
  d\sigma_{H^{(2)}} \equiv \left(\frac{\alphas(\muR)}{2\pi}\right)^{\!2} H^{(2)}(Q)\, d\sigma_{\rm LO} \,,
\end{equation}
where a summation over the $q{\bar q}$ and $gg$ partonic channels is left understood.

The NNLO coefficient $H^{(2)}$ is estimated by applying two independent approximations to both the numerator and denominator of Eq.~\eqref{eq:H}.
Effectively, this \textit{reweighting} procedure corresponds to a rescaling of the approximated two-loop
finite remainder by the exact Born amplitude, thus significantly improving the quality of the approximations. 

The first approach relies on a \textit{soft Higgs boson approximation} introduced for the first time in Ref.~\cite{Catani:2022mfv}. 
In this limit, the all-order $t \bar t H$ finite remainder satisfies the following factorisation formula,
\begin{equation}
  \label{eq:soft_Higgs}
	\M^{\mathrm{fin}}(\{p_i\},q; \muR, \muIR) \simeq F(\alphas(\muR), \muR/m_t)\,\frac{m_t}{v} \,\left( \frac{m_t}{p_3 \cdot q} + \frac{m_t}{p_4 \cdot q}\right) \M^{\mathrm{fin}}_{t \bar t}(\{p_i\}; \muR, \muIR) \,,
\end{equation}
where \mbox{$v=(\sqrt{2}G_\mu)^{-1/2}$} is the Higgs vacuum expectation value and $m_t$ the top-quark mass,
$p_3$ and $p_4$ are the four-momenta of the final-state top quarks, and $q$ is the momentum of the Higgs boson.
The perturbative function $F(\alphas(\muR), \muR/m_t)$ can be extracted by taking the soft limit of the heavy-quark scalar form factor,
whose explicit expression up to $\mathcal{O}(\alphas^3)$ is given in Ref.~\cite{Devoto:2024nhl}.
With $\M^{\mathrm{fin}}_{t \bar t}$ we denote the finite remainder of the scattering amplitude for $t\bar t$ production, which is known up to the two-loop order~\cite{Barnreuther:2013qvf}.
In Eq.~\eqref{eq:soft_Higgs}, the symbol $\simeq$ means that we have neglected contributions that are less singular than $1/q$ in the soft-Higgs limit \mbox{$q \to 0$}.

In the second approach, the two-loop coefficient $H^{(2)}$ is approximated in the high-energy limit ($m_t \ll Q$) of the top quarks, via the so-called \textit{massification} procedure \cite{Penin:2005eh,Mitov:2006xs,Becher:2007cu,Engel:2018fsb,Wang:2023qbf}.
Up to power corrections in $m_t/Q$, we can write 
\begin{equation}
	\M^{\mathrm{fin}}(\{p_i\},q; \mu) \simeq \boldsymbol{\mathcal{F}}_{[c]}\left(\alphas(\mu),\frac{\mu^2}{m_t^2}, \frac{\mu^2}{2 p_i \cdot p_j} \right)\,\M_{(m_t=0)}^{\mathrm{fin}}(\{p_i\},q; \mu) \,,
	\label{eq:massification}
\end{equation}
where $\mu=\muIR=\muR$. 
In Eq.~\eqref{eq:massification}, $\M_{(m_t=0)}^{\mathrm{fin}}$ denotes the finite remainder for the production of a Higgs boson plus four massless partons,
available up to the two-loop order in Ref.~\cite{Badger:2021ega,Badger:2024awe}, whereas \mbox{$\boldsymbol{\mathcal{F}}_{[c]}$} is a perturbative process-dependent
colour operator, whose expression is given in Ref.~\cite{Devoto:2024nhl}, and depends on the partonic channel $c=q,g$.  

The two-loop contribution is ultimately obtained by combining the results from the soft~\eqref{eq:soft_Higgs} and high-energy~\eqref{eq:massification} approximations through a weighted average.
Together with the estimate of the double-virtual contribution, a procedure was worked out to quantify the systematic error due to such an approximation.
More specifically, for each partonic channel, an error on $d\sigma_{H^{(2)}}$ is separately defined for the soft approximation and the massification approach. This error takes into account the relative discrepancy between the exact and approximated predictions at NLO as well as the effects on $d\sigma_{H^{(2)}}$ due to the variation of the subtraction scale $\muIR$, at which the approximations in Eqs.~\eqref{eq:soft_Higgs} and \eqref{eq:massification} are applied.
For further details on this procedure, we refer the reader to section 3 of Ref.~\cite{Devoto:2024nhl}.

The two-loop contribution in Eq.~(\ref{eq:sigma2}), computed as discussed above, is finally combined with the remaining NNLO terms, all evaluated exactly (including the one-loop squared contribution), within the {\sc Matrix} framework~\cite{Grazzini:2017mhc}.
In {\sc Matrix}, IR singularities are handled and cancelled via a process-independent implementation of the $q_T$-subtraction formalism extended to heavy-quark production, as mentioned above.
All NLO-like singularities are treated by dipole subtraction~\cite{Catani:1996jh,Catani:1996vz,Catani:2002hc,Kallweit:2017khh,Dittmaier:1999mb,Dittmaier:2008md,Gehrmann:2010ry,Schonherr:2017qcj}.
The required tree-level and one-loop amplitudes are obtained via
{\sc OpenLoops}~\cite{Cascioli:2011va, Buccioni:2017yxi,Buccioni:2019sur}
and {\sc Recola}~\cite{Actis:2012qn,Actis:2016mpe,Denner:2017wsf}.

The systematic error due to the approximation of the double-virtual contribution turns out to be sufficiently small to be subdominant with respect to 
the perturbative uncertainties at NNLO QCD. However, the uncertainty due to such an approximation cannot be neglected within 
the theory-uncertainty budget of the matched results presented in the following. 

\subsection{NNLL resummation in SCET}
\label{sec:theo-nnllb}
The approach to soft gluon emission resummation based on SCET is reviewed in this section.
The resummation framework relies on the factorisation of the partonic
cross section in the soft emission limit.
For the case of the $ pp \to t \bar{t} H$ production process,
the factorisation formula is derived following closely the same
procedure applied to threshold resummation in Drell-Yan \cite{Becher:2007ty}
(for a didactic introduction, see also \cite{Becher:2014oda})
and top-pair production \cite{Ahrens:2010zv}.
The case of the associated production of a top-pair
and a Higgs boson is considered in detail in
Refs.~\cite{Broggio:2015lya, Broggio:2016lfj}.
The closely related cases of $t \bar{t} W$ and $t \bar{t} Z$
production are addressed in Refs.~\cite{Broggio:2016zgg, Broggio:2017kzi}.
The interested reader can find a more
detailed discussion of the content of this section in
Refs.~\cite{Broggio:2016zgg, Broggio:2016lfj}.

In momentum space, the parameter that regulates the soft limit is $z = Q^2/\hat{s}$, where $Q$ is the invariant mass of the $t \bar{t} H$ final state and $\sqrt{\hat{s}}$ is the partonic centre of mass energy. In the soft limit, $z \to 1$, the total cross section factorizes as follows:
\begin{align}
\sigma\left(S, \mt, m_H \right) =& \frac{1}{2 S} \int_{\tau_{{\tiny min}}}^1 d \tau \int_{\tau}^1 \frac{dz}{\sqrt{z}} \sum_{ij} f\!\!f_{ij} \left( \frac{\tau}{z}, \mu \right) \nonumber \\
& \times \int d\mbox{PS}_{\tth}  \mbox{Tr} \left[ \mathbf{H}_{ij}\left(\{p \}, \mu \right) \mathbf{S}_{ij}\left( \frac{Q (1-z)}{\sqrt{z}}, \mu \right)\right] \, , \label{b:tot}
\end{align}
where $\sqrt{S}$ is the hadronic centre of mass energy,
\begin{equation}
\tau_{\mbox{{\tiny min}}} = \frac{\left(2 \mt + m_H \right)^2}{S} \, , \qquad \tau = \frac{Q^2}{S} \, ,
\end{equation}
and the symbol $\{p\}$
indicates the set of the momenta of the incoming partons $i$
and $j$ together with the top quark, anti-top quark and Higgs momenta.
The parton luminosity functions $f \! \!f_{ij}$
are defined as the convolutions of the PDFs of the $i$ and $j$ partons.
In the case of $\tth$ production,
the two channels contributing to the factorisation formula in the
soft limit are the quark annihilation channel and the gluon fusion channel.
The phase space measure for the
$\tth$ final state is indicated by $d \mbox{PS}_{\tth}$.
The trace $\mbox{Tr} \left[\mathbf{H} \mathbf{S}\right]$
is proportional to the spin
and colour averaged partonic matrix elements for the
$\tth + X_s$ production process,
where $X_s$ indicates the unobserved soft gluons in the final state.
The hard functions $\mathbf{H}_{ij}$ are matrices in colour space
and can be calculated starting from the colour decomposed amplitudes
for the virtual corrections to the partonic tree-level
$\tth$ production diagrams.
Also, the soft functions $\mathbf{S}_{ij}$ are matrices in colour space.
The soft functions can be obtained from the colour
decomposed real emission amplitudes evaluated in the soft limit $z \to 1$.

To perform soft-gluon resummation it is convenient to rewrite the  total cross section (\ref{b:tot}) in terms of the Mellin transformed partonic differential cross section as
\begin{align}
\sigma\left(S, \mt, m_H \right) = \frac{1}{2 S} \int_{\tau_{{\tiny min}}}^1 \frac{d \tau}{\tau} \frac{1}{2 \pi i} \int_{c -i \infty}^{c+ i \infty} \frac{dN}{\tau^N} \sum_{ij} \widetilde{f\!\!f}_{ij} \left( N, \mu \right)  \int d\mbox{PS}_{\tth}  d\widetilde{\hat{\sigma}}_{ij} \left(N,\mu \right)\, , \label{b:totMellin}
\end{align}
where the tilde indicates the Mellin transform of the luminosity function and of the partonic differential cross section. In particular, the Mellin transform of the partonic differential cross section, a.k.a. the hard scattering kernel, can be written as 
\begin{equation}
d\widetilde{\hat{\sigma}}_{ij} \left(N,\mu \right) = \mbox{Tr} \left[ \mathbf{H}_{ij} \left( \{p\}, \mu \right)\widetilde{\mathbf{s}}_{ij} \left(\ln{\frac{Q^2}{\bar{N}^2 \mu^2}}, \mu \right) \right] \, . \label{b:partonic}
\end{equation} 
Since the soft limit $z \to 1$ corresponds to the limit $N \to \infty$ in Mellin space,
terms suppressed by powers of $1/N$ were neglected in Eq.~(\ref{b:totMellin}). The quantity $\bar{N}$ that appears in the Mellin transform of the soft function, $\widetilde{\mathbf{s}}$, is defined by the relation $\bar{N} = N e^{\gamma_E}$, where the Euler constant is $\gamma_E \approx 0.577216 \cdots$.

The hard $\mathbf{H}$ and soft $\widetilde{\mathbf{s}}$ functions,
appearing in the hard scattering kernels in Eq.~(\ref{b:partonic}),
can be evaluated in fixed-order perturbation theory at values of the scale
$\mu$ at which they
are free from numerically large logarithms.
The scale chosen for the evaluation of the hard function is indicated by $\mu_h$
and that for the soft function is indicated by $\mu_s$.
Treating $\mu_s / \mu_h \ll 1$, there is thus no common value $\mu$
which can be made to eliminate large logarithms in both functions
simultaneously.
This problem is circumvented by deriving
renormalisation-group equations (RGEs) that
can be solved to evolve the hard
and soft functions from their natural scales $\mu_h$
and $\mu_s$ to a common factorisation scale $\muF$
at which the PDFs are evaluated.
Formally, the result of this operation is \cite{Broggio:2016lfj}
\begin{align}
d\widetilde{\hat{\sigma}}_{ij} \left(N,\muF \right) =& \mbox{Tr} \left[ \widetilde{\mathbf{U}}_{ij} \left(\bar{N}, \{p\}, \muF, \mu_h, \mu_s \right) \mathbf{H}_{ij} \left( \{p\}, \mu_h \right) \widetilde{\mathbf{U}}^\dagger_{ij} \left(\bar{N}, \{p\}, \muF, \mu_h, \mu_s \right)\right. \nonumber \\
 &  \,\,\,\,\quad \times \left. \widetilde{\mathbf{s}}_{ij} \left(\ln{\frac{Q^2}{\bar{N}^2 \mu_s^2}}, \mu_s \right)\right] \, . \label{b:kernelmellin}
\end{align}
The evolution factors $ \widetilde{\mathbf{U}}$, which depend on the partonic channel, resum large logarithmic corrections depending on the ratio $\mu_s/\mu_h$. They can be expressed as \cite{Broggio:2016zgg}
\begin{align}
\widetilde{\mathbf{U}}_{ij} \left(\bar{N}, \{p\}, \muF, \mu_h, \mu_s \right) =& \exp \left\{ \frac{4 \pi}{\alpha_s (\mu_h)} g_1 \left(\lambda_s, \lambda_f \right) + g_2 \left(\lambda_s, \lambda_f \right) +  \frac{\alpha_s (\mu_h)}{4 \pi} g_3 \left(\lambda_s, \lambda_f \right) + \cdots \right\} \nonumber \\
& \times \mathbf{u}_{ij} \left(\{p \}, \mu_h, \mu_s \right) \, , \label{b:evolutionU}
\end{align}
where $\mathbf{u}$ is the non-diagonal part of the evolution matrix, and 
\begin{equation}
\label{eq:lambda_SCET}
\lambda_s \equiv \frac{\alpha_s(\mu_h)}{2 \pi} \beta_0 \ln{\frac{\mu_h}{\mu_s}} \, , \qquad  
\lambda_f \equiv \frac{\alpha_s(\mu_h)}{2 \pi} \beta_0 \ln{\frac{\mu_h}{\muF}} \, .
\end{equation}
The functions $g_i$ depend on the cusp and PDFs anomalous dimensions. The function $g_1$ is referred to as the leading logarithmic (LL) function, the function $g_2$ is known as the next-to-leading logarithmic (NLL) function, etc.\footnote{Explicit results can be found in Appendix C.1 of \cite{Czakon:2018nun}, where the functions $g_i$ in Eq.~(\ref{b:evolutionU}) are denoted instead by $g_i^m$.}  

At all orders in perturbation theory, the l.h.s.\ of Eq.~(\ref{b:kernelmellin}) does not depend on the choice of the hard and soft scales, $\mu_h$ and $\mu_s$. However, in practice the hard and soft functions can only be evaluated up to some finite order in perturbation theory. This fact introduces in any numerical evaluation of Eq.~(\ref{b:kernelmellin}) a residual dependence on the choice of $\mu_h$ and $\mu_s$. The hard and soft scales are chosen such that $\mu_h \sim Q$ and $\mu_s \sim Q/\bar{N}$. 
While this choice of soft scale allows all large logarithms involving the Mellin parameter $N$
to be resummed, when integrating over 
$N$ as in Eq.~(\ref{b:totMellin}) to obtain the physical cross section, one faces the well-known problem of a branch cut for large values of $N$ in the hard scattering kernel, which is related to the existence of the Landau pole in $\alpha_s$. This issue is taken care of by adopting the \emph{Minimal Prescription} introduced in Ref.~\cite{Catani:1996yz}.

The resummed formulas include certain towers of logarithms to all orders in perturbation theory, but neglect contributions that are subleading in the soft limit. These subleading corrections can be added back in fixed-order perturbation theory through a matching procedure. For the NNLO+NNLL result in $ttH$ production, this matching procedure is implemented by evaluating the hadronic differential cross section according to
\begin{align}
 \label{b:NNLOpNNLL}
 d\sigma_{ttH}^{\mathrm{NNLO+NNLL}} =d\sigma_{ttH}^{\mathrm{NNLL}}
+\bigg(d\sigma_{ttH}^{\mathrm{NNLO}}-
d\sigma_{ttH}^{\mathrm{NNLL}}\Big|_{\substack{\text{NNLO} \\ \text{expansion} }}\bigg) \, ,
\end{align}
where the third term above is the NNLO expansion of the NNLL resummation formula, which is obtained  
by treating logarithms of scale ratios as ${\cal O}(1)$ quantities and re-expanding
the NNLL result to the second relative order in $\alpha_s(\muF)$. The most non-trivial contribution is the second-order correction, 
 which is derived in complete analogy to the  top-pair production case \cite{Czakon:2018nun} and involves
 a term in the Mellin-transformed partonic cross section which reads
\begin{align}
d\widetilde{\hat{\sigma}}_{ij}^{(2)} \left(N,\muF \right) =& \mbox{Tr} \left[ \mathbf{H}_{ij}^{(2)} \left( \muF \right)\tilde{\mathbf{s}}_{ij}^{(0)}   \left( \muF \right)+ \mathbf{H}_{ij}^{(1)} \left( \muF \right)\tilde{\mathbf{s}}_{ij}^{(1)}   \left( \muF \right)+ \mathbf{H}_{ij}^{(0)} \left( \muF \right)\tilde{\mathbf{s}}_{ij}^{(2)}   \left( \muF \right) \right] \nonumber \\
&- \mbox{Tr} \left[ \mathbf{H}_{ij}^{(2)} \left( \mu_h \right)\tilde{\mathbf{s}}_{ij}^{(0)}   \left( \mu_s \right)+ \mathbf{H}_{ij}^{(1)} \left( \mu_h \right)\tilde{\mathbf{s}}_{ij}^{(1)}   \left( \mu_s \right)+ \mathbf{H}_{ij}^{(0)} \left( \mu_h \right)\tilde{\mathbf{s}}_{ij}^{(2)}   \left( \mu_s \right) \right] \, , \label{b:NNLOreexpansion}
\end{align}
where the superscript $(n)$, $n=0,1,2$, indicates the order in $\alpha_s$ at which the various contributions are evaluated.
The formula in Eq.~(\ref{b:NNLOpNNLL}) for the NNLO+NNLL cross section is such that the first term takes 
into account the all-order soft-gluon resummation to NNLL, while the combination of the terms in parenthesis adds to it the subleading pieces in the soft limit to NNLO in fixed order. 
 
\subsubsection{%
    Introduction of the renormalisation scale
    and study of residual scale dependence%
}

The residual scale uncertainty
affecting the phenomenological predictions presented in
Refs.~\cite{Broggio:2016lfj, Broggio:2019ewu}
was assessed through the conventional procedure in SCET,
namely by independently varying the hard, soft
and factorisation scales present in the resummation formula by factors of $2$
and $1 / 2$ with respect to their default choices.
Subsequently, the three types of scale variation were then added in
quadrature in order to quote a total scale uncertainty,
as detailed in section 3 of Ref.~\cite{Broggio:2016lfj}
and section 3.2 of Ref.~\cite{Broggio:2019ewu}.

While the approach adopted in
Refs.~\cite{Broggio:2016lfj, Broggio:2019ewu} is sound
and reasonably conservative, it makes a direct comparison with the results in
Refs.~\cite{Kulesza:2015vda, Kulesza:2017ukk} somewhat cumbersome.
This is due to the fact that the results in
Refs.~\cite{Kulesza:2015vda, Kulesza:2017ukk}
allow to vary separately the factorisation and renormalisation scales,
which are set equal in Refs.~\cite{Broggio:2016lfj, Broggio:2019ewu},
while in Refs.~\cite{Kulesza:2015vda, Kulesza:2017ukk} the soft
and hard scales are kept fixed.
Moreover, the uncertainties
in fixed-order NNLO calculations are typically evaluated
through a 7-point variation of the factorisation and
renormalisation scales,
so in order to compare such calculations with NNLO+NNLL results one must retain
distinct factorisation
and renormalisation scales also in the resummed part of the calculation.

For these reasons, the $\tth$
cross-section predictions evaluated in this work through the method of
Refs.~\cite{Broggio:2016lfj, Broggio:2019ewu}
are obtained after introducing the renormalisation scale
$\muR$ in the resummation formula.
This is done by eliminating $\alpha_s(\mu_h)$
in favour of $\alpha_s(\muR)$ by means of the relation
\begin{align}
\alpha_s(\mu_h) = & \frac{\alpha_s (\muR)}{X} \left[ 1- \frac{\alpha_s(\muR)}{4 \pi}  \frac{\beta_1}{\beta_0} \frac{\ln X}{X} + \left( \frac{\alpha^2_s(\muR)}{4 \pi} \right)^2 \left( \frac{\beta^2_1}{\beta^2_0} \frac{\ln^2{X} - \ln{X} -1 + X}{X^2} \right. \right.
\nonumber \\
&\left. \left.+ \frac{\beta_2}{\beta_0} \frac{1-X}{X}\right) + \cdots \right] \, , 
\end{align}
where 
\begin{equation}
X = 1- \frac{\alpha_s(\muR)}{2 \pi} \beta_0 \ln{\frac{\muR}{\mu_h}} \, . 
\end{equation}
Once the scale $\muR$ has been introduced, the resummed partonic cross section is re-expanded in powers of $\alpha_s(\muR)$,
treating logarithms of any two scale ratios as ${\cal O}(1)$.   

It is then necessary to specify how the residual scale uncertainty affecting the predictions is obtained. In order to make comparisons with the results discussed in Refs.~\cite{ Kulesza:2018tqz,Broggio:2019ewu,Kulesza:2020nfh}, the soft scale $\mu_s$ was set equal to $Q/\bar{N}$ irrespectively from the choice made for the other scales. Three different choices were made for the central values 
$\mu_0$ given to $\muF$ and $\muR$: $\mu_0$ was set equal to 
a) $\mu_0 = Q/2$,  b) $\mu_0 = H_T/2$, or  c) $\mu_0 = \mt+m_H/2$.  


For each choice of $\mu_0$,
the factorisation and renormalisation scales were varied independently
by employing the usual 7-point method.
In addition, a value for $\mu_h$ must be chosen.
For each choice of $\muF$ and $\muR$,
the cross section was evaluated both with $\mu_h = \muF$
and with $\mu_h = \muR$.
In summary, for each $\mu_0$
the total cross section at NNLO+NNLL was evaluated for 11 scale choices:
\begin{align}
\label{eq:SCET_11}
	(\mu_F/\mu_0,\mu_R/\mu_0,\mu_h/\mu_0) \in \{&(1,1,1),(2 ,1,2),(2,1,1),(1/2,1,1/2),(1/2,1,1),(1,2,1),\,\nonumber\\
	& (1,2,2),(1,1/2,1/2),(1,1/2,1),(2,2,2),(1/2,1/2,1/2)\} \, .
\end{align}
Finally, the scale uncertainty affecting the cross section is
determined by taking the envelope of these 11 scale choices.\footnote{While $\mu_s  = Q/\bar{N}$ is kept fixed across these 11 scale choices, we verified that varying it up and down by a factor of 2 for $\mu_h = \mu_F = \mu_R = \mu_0$ does not change the uncertainty envelopes for results obtained in this paper.}
%
\subsection{NNLL resummation in dQCD}
\label{sec:theo-nnllk}
\newcommand{\tosv}{{\scriptscriptstyle \to}}

In this section,
we describe the calculations of the NNLO+NNLL cross section for the process
$pp \to \tth$, carried out in dQCD.
In this formalism, the resummation of large logarithmic corrections
in the threshold limit can be achieved
either by direct diagrammatic analysis~\cite{Catani:1989ne}
or all-order factorisation properties
of partonic cross sections~\cite{Sterman:1986aj}.
For processes with four or more coloured legs, which is the case here,
the non-trivial colour flow needs to be accounted for~\cite{Bonciani:1998vc, Kidonakis:1997gm}.
The first application of threshold resummation to calculate the
$\tth$ cross section was carried out in Ref.~\cite{Kulesza:2015vda},
where the process was considered at the absolute production threshold limit.
Here we briefly review the calculations presented in Ref.~\cite{Kulesza:2017ukk},
where the resummed $\tth$
cross section was obtained at the NNLL accuracy using the threshold definition
with respect to the invariant mass of the final state system.
The same formalism has also been employed to obtain the NNLL predictions for the
$t \bar t Z$ and
$t \bar t W$ production processes~\cite{Kulesza:2018tqz, Kulesza:2020nfh}.

The resummation of logarithmic corrections which become large close to the production threshold, i.e. when the invariant mass $Q^2$ of the $\tth$ system approaches the partonic centre of mass energy $\hat s$,  takes place in the space of Mellin moments $N$. At the partonic level, the Mellin transform of $ d \hat \si / dQ^2 $  reads 
\begin{equation}
 \frac{\text{d} \tilde {\hat \sigma}_{ij \tosv \tth}}{\text{d} Q^2} (N,Q^2,  \{m^2\},\muR^2,\muF^2)= \int_0^1 \text{d} \hat \rho \; \hat \rho^{N-1} \frac{\text{d} \hat \sigma_{ij \rightarrow t\bar t H}}{\text{d} Q^2} (\hat \rho,Q^2, \{m^2\},\muR^2,\muF^2),
\end{equation}
where $\hat \rho = Q^2 / \hat s$,
$\{m^2\}$ stands for all masses entering the calculations
and $i, j$ denote two initial-state coloured partons.
The cross section factorises into a product of
Mellin-transformed functions of the coupling constant
and the ratios of the scales,
\begin{eqnarray}
\label{eq:res:fact:dqcd}
&\frac{d\tilde{\hat \si}^{\rm (NNLL)}_{ij\tosv \tth}}{dQ^2} &\!\!\!(N, Q^2,\{m^2\},\muR^2,\muF^2) = {\mathrm{Tr}}\left[ \mathbf{H}_R (Q^2, \{m^2\},\muR^2, \muF^2)  \right. \nonumber\\
&& \times \left. \mathbf{\bar{U}}_R(N+1, Q^2,\{m^2\}, \muR^2) 
\mathbf{\tilde S}_R (N+1, Q^2, \{m^2\})\, \mathbf{{U}}_R(N+1, Q^2,\{m^2\}, \muR^2 )  \right] \nonumber \\
&&\times\,
\Delta^i(N+1, Q^2,\muR^2, \muF^2 ) \; \Delta^j(N+1, Q^2,\muR^2, \muF^2 ),
\end{eqnarray}
where $\mathbf{H}_R$, $\mathbf{\bar{U}}_R$,
$\mathbf{{U}}_R$ and $\mathbf{\tilde S}_R$
are matrices in colour space over which the trace is taken.
The jet functions $\Delta^i$ account for (soft-)collinear
logarithmic contributions from the initial state partons
and are well known at NNLL~\cite{Catani:1996yz, Catani:2003zt}.
The term $\mathbf{\bar{U}}_R \, \mathbf{\tilde S}_R \, \mathbf{{U}}_R$
originates from a solution of the
renormalisation group equation for the soft function
and consists of the evolution matrices $\mathbf{\bar{U}}_R$,
$\mathbf{{U}}_R$, as well as the function $\mathbf{\tilde S}_R$
which plays the role of a boundary
condition of the renormalisation group equation.
The evolution matrices are given by
(reverse in the case of $\mathbf{\bar{U}}_R$)
path-ordered exponentials of the soft anomalous dimension matrix {%
    $
        \mathbf{\bar \Gamma}_{ij \tosv \tth}(\alphas)
        = \left(\frac{\alphas}{\pi}\right)
        \mathbf{\bar \Gamma}^{(1)}_{ij \tosv \tth}
        + \left(\frac{\alphas}{\pi}\right)^2
        \mathbf{\bar \Gamma}^{(2)}_{ij \tosv \tth}
        + \ldots
    $%
} which is obtained by subtracting the contributions
already taken into account in $\Delta^i \Delta^j$
from the full soft anomalous dimension for the process $ij \to \tth$.
At NLL, the path-ordered exponentials collapse to standard
exponential factors in the colour space
$\mathbf R$ where $\mathbf \Gamma^{(1)}_R$ is diagonal.
At NNLL,
the path-ordered exponentials are eliminated by treating $\mathbf{U}_R$
and $\mathbf{\bar{U}}_R$ perturbatively
\beq
\label{eq:UR}
\mathbf{U}_R(N,Q^2,\{m^2\}, \muR^2 )=\left(\mathbf{1}+\frac{\alphas(\muR^2)}{\pi[1-2 \alphas(\muR^2) b_0 \log N]}\mathbf{K}\right)
\left[e^{\,g_s(N)\overrightarrow{\gamma}^{(1)}} \right]_{D}
\left(\mathbf{1}-\frac{\alphas(\muR^2)}{\pi}\, \mathbf{K}\right),
\eeq
where
$\overrightarrow{\gamma}^{(1)}$ is a vector of  $\pmb{\Gamma}^{(1)}_{ij\to klB}$ eigenvalues and subscript $D$ indicates a diagonal matrix. Furthermore, $\mathbf{K}_{IJ}=\delta_{IJ}{\gamma}^{\left(1\right)}_{I}\frac{b_1}{2b_0^2}-\frac{\left(\pmb{\Gamma}^{(2)}_R\right)_{IJ}}{2\pi b_0+\gamma^{\left(1\right)}_{I}-\gamma^{\left(1\right)}_{J}}\,$ with $b_0$ and $b_1$ denoting the first two $\beta_{\mathrm{QCD}}$ coefficients and
\begin{eqnarray}
g_s (N)= \frac{1}{2 \pi b_0} \left\{  \log(1-2\lambda) + \alphas(\muR^2) \left[   \frac{b_1}{b_0} \frac{ \log(1-2\lambda)}{ 1-2\lambda} - 2 \gamma_{\rm E} b_0  \frac{2\lambda}{1-2\lambda} \right. \right. \nonumber \\
\left. \left.  
+ \, b_0 \log\left( \frac{Q^2}{\muR^2} \right) \frac{2\lambda}{1-2\lambda} \right] \right\}
\label{eq:gs_dQCD}
\end{eqnarray}
with $\lambda = \alphas(\muR^2) b_0 \log N$.
The remaining function in Eq.~(\ref{eq:res:fact:dqcd}),
$\mathbf{H}_R$, contains information on the hard off-shell
dynamics and collects contributions non-logarithmic in
$N$ which are projected on the $\mathbf R$ colour basis.
At NNLL,
the ${\cal O}(\alphas)$ terms in the perturbative expansion of $\mathbf{H}_R$
and $\mathbf{\tilde S}_R$, as well as $\mathbf{\Gamma}^{(2)}_R$ are needed.
While the latter is known analytically~\cite{Ferroglia:2009ep},
the contributions of the other two functions need to be determined.
In particular,
the virtual corrections which enter $\mathbf{H}^{\mathrm{(1)}}_{R}$
are extracted numerically from the NLO calculations by
{\tt MadGraph5\_aMC@NLO}~\cite{Alwall:2014hca}.

The threshold-resummed NNLL cross sections are then matched to the NNLO predictions of Ref.~\cite{Devoto:2024nhl} according to
\beq
 {d\sigma^{\rm (NNLO+NNLL)}_{pp \tosv \tth} \over dQ^2}  =
\frac{d\sigma^{\rm (NNLO)}_{pp \tosv \tth}}{dQ^2} 
+ \frac{d \sigma^{\rm (res-exp)}_{p p \tosv \tth}}{dQ^2}
\eeq
 with
\beqar
\label{invmel}
&& \frac{d \sigma^{\rm
  (res-exp)}_{p p \tosv \tth}}{dQ^2} (Q^2,\{m^2\},\muR^2, \muF^2) \! =   \sum_{i,j}\,
\int_{\sf C}\,\frac{dN}{2\pi
  i} \; \rho^{-N} f^{(N+1)} _{i/h{_1}} (\muF^2) \, f^{(N+1)} _{j/h_{2}} (\muF^2) \nn \\ 
&& \! \times\!\! \left[ 
\frac{d \tilde{\hat \si}^{\rm (NNLL)}_{ij\tosv \tth}}{dQ^2} (N,Q^2,\{m^2\},\muR^2, \muF^2)
\!-\!  \frac{d \tilde{\hat \si}^{\rm (NNLL)}_{ij\tosv \tth}}{dQ^2} (N,Q^2,\{m^2\},\muR^2, \muF^2)
{ \left. \right|}_{\scriptscriptstyle({\rm NNLO})}\, \!\! \right]\!\!, 
\eeqar
where $f_{i / h}(x, \muF^2)$ are moments of the parton
distribution functions and $
    d \hat \si^{\rm(res)}_{ij \tosv \tth} / dQ^2 \left. \right|_{
        \scriptscriptstyle({\rm NNLO})
    }
$
represents the perturbative expansion
of the resummed cross section truncated at NNLO.
The inverse Mellin transform (\ref{invmel}) is evaluated numerically
according to the ``Minimal Prescription'' ~\cite{Catani:1996yz}
along a contour ${\sf C}$ in the complex-$N$ space.
For more information on the theoretical framework,
we refer the reader to Refs.~\cite{%
    Kulesza:2017ukk, Kulesza:2018tqz, Kulesza:2020nfh%
}.

\subsection{Comparison of the two resummation approaches}
\label{sec:theo-nnllcomp}
Although the formulas for the resummed $\tth$ cross sections presented in the previous two subsections are derived in conceptually distinct frameworks and look quite different at first glance,
the two resummation formalisms,
describing the same physics in the soft gluon emission limit,
are theoretically equivalent.
In particular,
both make use of factorisation in the soft limit along with RG-improved
perturbation theory to resum logarithmic corrections,
and if the formulas were evaluated to infinite
logarithmic accuracy, they would agree exactly.
It needs to be stressed, however,
that the derivation of the two formulas is performed in very different ways.
In the case of dQCD, the formalism is derived directly from the properties of
scattering amplitudes in full QCD,
while in SCET, effective field theory techniques are used in intermediate steps.
The two distinct theoretical frameworks lead naturally to different organisations of the resummed
expressions,
so that when evaluated at a fixed logarithmic accuracy, the analytic
and numerical
results are no longer the same.  Before exploring numerical results, we first highlight some 
of the salient differences in the analytic expressions. 

In the context of the present work, a particularly noticeable difference between the two 
formalisms is the set of scales that are allowed to vary
and the parametric counting of logarithms underlying
RG-improved perturbation theory and thus resummation.
In dQCD,
one can vary the two scales $\mu_F$ and $\mu_R$, and
$\lambda = \alpha_s(\muR^2 ) \, b_0 \ln N$ is considered an order one parameter.  In SCET, on 
the other hand, as explained in section~\ref{sec:theo-nnllb}, the set of scales
$\mu_i \in \{\muF, \muR, \mu_h\}$ is allowed to vary,  and the ratio 
of any of these scales with each other or with $\mu_s = Q/\bar{N}$ is considered a large 
logarithm.  As a result, while in dQCD expansion coefficients 
such as Eq.~(\ref{eq:gs_dQCD}) depend only on $\lambda$, expansion coefficients
in SCET can depend on several order-one parameters -- for example, the expansion
coefficients $g_i$ in  Eq.~(\ref{b:evolutionU}) depend on $\lambda_s$ and $\lambda_f$ 
in Eq.~(\ref{eq:lambda_SCET}), for the case $\mu_h= \mu_R$. Other notable differences include
the fact that the exponential factors in dQCD (but not in SCET) are chosen to vanish in the limit $\lambda \to 0$ (see e.g. Eq.~(\ref{eq:gs_dQCD})), which is achieved by re-expanding the $\lambda$-independent terms and absorbing them into the hard function,\footnote{Note that due to this and other similar manipulations,  the dQCD hard functions $\mathbf{H}_R$ are {\rm not} identical to the SCET hard functions $\mathbf{H}$. For the same reason, contributions involving the factor $\gamma_E$ differ by terms considered N$^3$LL and higher orders in dQCD.} while in SCET (but not dQCD) the approximation  $\exp (\alphas g_3) \approx 1 + \alpha_s g_3$ is used.  

We have checked analytically that when exactly the same implementation
of RG-improved perturbation theory is used,
the SCET and dQCD formulas agree at NNLL.
For numerical evaluations we have retained the differences
in the two setups as outlined above,
so that the SCET formulas contain some corrections that are considered N$^3$LL
and higher order (and thus not necessarily included) in the dQCD formalism
and vice versa.
The numerical differences between the predictions obtained within the two formalisms can 
then be used as an additional handle on theoretical uncertainties in the soft gluon resummation formulas, beyond those estimated through scale variations in either approach alone. In particular, these differences can be seen as an indicator of the size of subleading terms beyond the formal accuracy of resummation, i.e. N$^3$LL and higher. 

The comparison of NNLO+NNLL results for the total cross section in SCET and dQCD is shown on the left-hand side of figure~\ref{fig:tth_comparisons}. The results are shown for three different parametric choices
for the default values of $\muF$ and  $\muR$:
\begin{itemize}
\item $\muF =\muR = m_t+m_H/2$
\item $\muF =\muR =H_T/2$
\item $\muF =\muR   \equiv Q/2$ ($Q\equiv M_{ttH}$) 
\end{itemize}
Scale uncertainties in SCET  are obtained by evaluating the cross section for the 11 different values for
$\muF,\muR$ and $\mu_h$ listed in Eq.~(\ref{eq:SCET_11}), while those in dQCD are evaluated using the standard 7-point 
method.  We observe that the NNLO+NNLL results agree remarkably well, with the central values differing only by a few permille. 

In order to take the small differences between the two approaches as an additional theoretical uncertainty, 
we combine dQCD and SCET results by averaging the central values and taking the envelope of the uncertainty bands.
The result of this combination is shown on the
right-hand side of figure~\ref{fig:tth_comparisons}, where we also display
the NNLO QCD results with uncertainties obtained via the 7-point method.
Comparing the two sets of results, one sees that combined NNLO+NNLL
predictions have not only smaller scale variation
errors but are also more stable with respect to the choice
of the default values of
$\muF = \muR$ than the NNLO results.
One also sees that the resummation effects are smallest for the default choice
$\muF = \muR = m_t+m_H/2$, which gives an additional motivation for using this choice when compiling final results
in section~\ref{sec:results} (%
    apart from the fact that this is the only physical scale available for total
    cross section,
    for which the values of dynamic scales have been integrated over%
).

\begin{figure}
    \centering
\includegraphics[width=.49\textwidth]{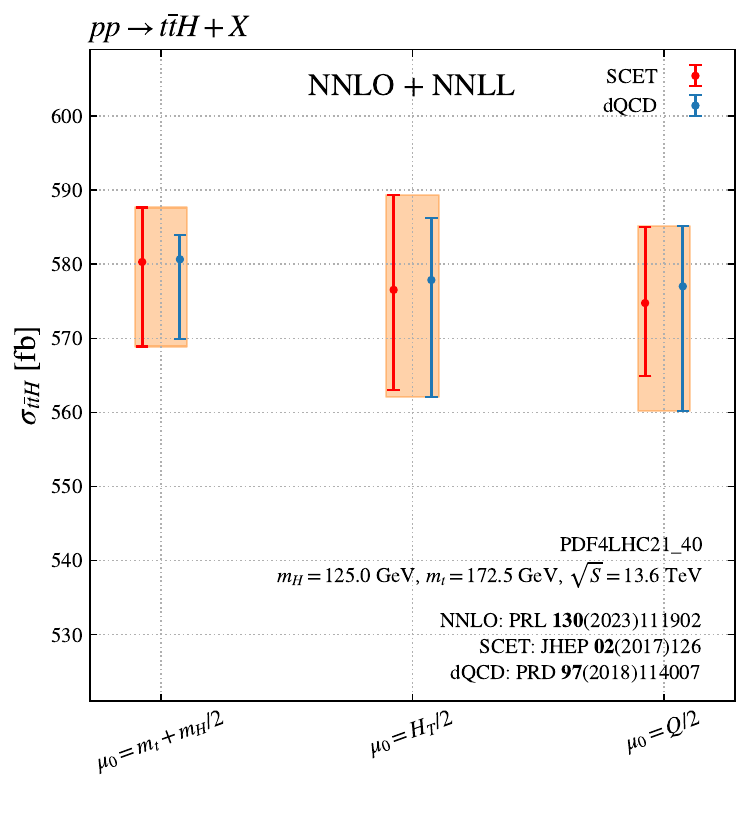}
\includegraphics[width=.49\textwidth]{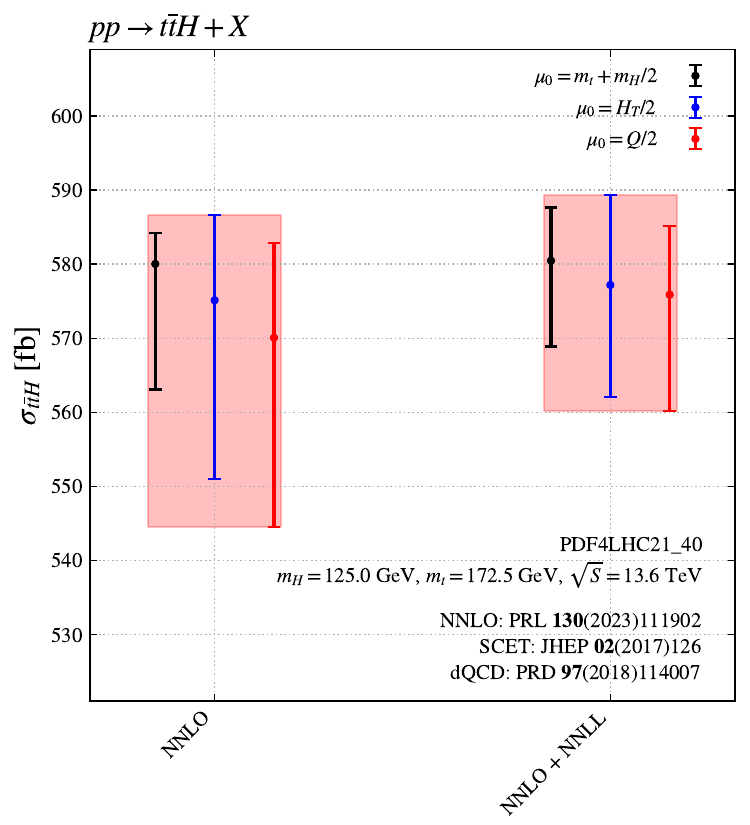}
    \caption{%
        Left: comparison between NNLO+NNLL results in dQCD
        and SCET for three parametrically different choices of the default
        scales.
        Right: comparison of the combined NNLO+NNLL results with NNLO
        for the same three sets of scales.
        No EW corrections are included.
        See the text for additional explanations
        on the estimation of the uncertainties. %
    }
    \label{fig:tth_comparisons}
\end{figure}
%
\subsection{NLO EW and photon-induced contributions}
\label{sec:theo-ew}

In this section, we briefly outline the structure of the contributions that enter the EW corrections to $t\bar t  H$. The notation is the same as used in 
Refs.~\cite{Frixione:2014qaa,Frixione:2015zaa,Czakon:2017wor,Frederix:2018nkq,Broggio:2019ewu}.

A given observable $\Sigma^{t\bar tH}$ for the process $pp \to t\bar tH(+X)$ can be simultaneously expanded in the QCD and EW couplings as:
\noindent
\begin{equation}
\Sigma^{t\bar tH}(\alpha_s,\alpha) = \sum_{m+n\geq 3} \alpha_s^m \alpha^n \Sigma_{m+n,n}\, .
\end{equation}
\noindent
The LO ($m+n=3$), NLO ($m+n=4$) and NNLO ($m+n=5$) contributions correspond  therefore to 
\begin{align}
\Sigma^{t\bar tH}_{\rm LO}(\alpha_s,\alpha) &= \alpha_s^2 \alpha \Sigma_{3,1} + \alpha_s \alpha^2 \Sigma_{3,2} + \alpha^3 \Sigma_{3,3} \nonumber\\
 &\equiv \Sigma_{\rm LO,1} + \Sigma_{\rm LO,2} + \Sigma_{\rm LO,3}\, ,  \nonumber\\
\Sigma^{t\bar tH}_{\rm NLO}(\alpha_s,\alpha) &= \alpha_s^3 \alpha \Sigma_{4,1} + \alpha_s^2 \alpha^2 \Sigma_{4,2} + \alpha_s \alpha^3 \Sigma_{4,3} + \alpha^4 \Sigma_{4,4} \nonumber\\
 &\equiv \Sigma_{\rm NLO,1} + \Sigma_{\rm NLO,2} + \Sigma_{\rm NLO,3} + \Sigma_{\rm NLO,4}\, ,  \nonumber\\
\Sigma^{t\bar tH}_{\rm NNLO}(\alpha_s,\alpha) &= \alpha_s^4 \alpha \Sigma_{5,1} + \alpha_s^3 \alpha^2 \Sigma_{5,2} + \alpha_s^2 \alpha^3 \Sigma_{5,3} + \alphas \alpha^4 \Sigma_{5,4} +  \alpha^5 \Sigma_{5,5} \nonumber\\
 &\equiv \Sigma_{\rm NNLO,1} + \Sigma_{\rm NNLO,2} + \Sigma_{\rm NNLO,3} + \Sigma_{\rm NNLO,4} + \Sigma_{\rm NNLO,5} \;.
\label{eq:blobs}
\end{align}
\noindent
The contributions $\Sigma_{\rm LO, 1}$,
$\Sigma_{\rm NLO, 1}$, and $\Sigma_{\rm NNLO, 1}$
are usually referred to as the LO contribution to the $t \bar t H$
cross section,
and its NLO and NNLO corrections in QCD; the quantity $\Sigma_{\rm NLO, 2}$
is usually referred to as the NLO EW corrections.
Finally, a prediction
including all LO and
NLO contributions is said to be computed at complete-NLO accuracy.
We will neglect NNLO contributions different than $\Sigma_{\rm NNLO, 1}$.

LO and NLO contributions different from $\Sigma_{\rm LO,1}$ and $\Sigma_{\rm NLO,1}$ can involve partonic processes with at least one photon in the initial state 
and therefore depend on the photon PDF. The dominant contribution originates from the process 
$g \gamma  \to t\bar t H$,\footnote{See Ref.~\cite{Pagani:2016caq} for an analogous and more detailed discussion for the case of $t \bar t$ production.} 
which enters both LO and NLO. However, also $ q\gamma$ and $\gamma\gamma$ initial states are possible. The quantities $\Sigma_{\rm NLO~EW}$,  $\Sigma_{\rm NLO,3}$ and $ \Sigma_{\rm NLO,4}$ receive contributions from the $q\gamma\to t\bar t H q$ processes, while the $\gamma \gamma$ initial state contributes to $\Sigma_{\rm LO,3}$, via $\gamma\gamma\to t\bar t H $, to $\Sigma_{\rm NLO,3}$, via $\gamma\gamma\to t\bar t H g$, and to $ \Sigma_{\rm NLO,4}$, via $\gamma\gamma\to t\bar t H \gamma$.

\subsection{Naming convention and construction of the state-of-the-art predictions}
\label{sec:names}
Having discussed all the theoretical ingredients entering the $t\bar t H$ cross section, we now introduce the naming convention for the quantities 
presented in the remainder of this work. We focus on the total cross section $\sigma$ and, owing to the dominance of QCD-type effects, we call
\begin{eqnarray}
    \sigma_{\rm LO} &\equiv&  \sigma_{\rm LO,1}\,,\\
    \sigma_{\rm NLO} &\equiv&  \sigma_{\rm LO,1} +\sigma_{\rm NLO,1}  \,,\\
    \sigma_{\rm NNLO} &\equiv&  \sigma_{\rm LO,1} +\sigma_{\rm NLO,1} +\sigma_{\rm NNLO,1}   \,.
\end{eqnarray}
These fixed-order predictions, in particular $\sigma_{\rm NNLO}$, can be supplemented with soft-gluon resummation (we consider only NNLL accuracy for simplicity). If we call 
$\sigma_{\rm NNLL}^{\rm SCET}$ ($ \sigma_{\rm NNLL}^{\rm dQCD}$) the purely resummed prediction at NNLL accuracy obtained with SCET (dQCD), the corresponding matched predictions are then obtained by adding these quantities to $\sigma_{\rm NNLO}$, and removing the double counting,
\begin{eqnarray}
    \sigma_{\rm NNLO+NNLL}^{\rm SCET}&=& \sigma_{\rm NNLO}+ \sigma_{\rm NNLL}^{\rm SCET} - \left.\sigma_{\rm NNLL}^{\rm SCET}\right|_{\alphas^2}\,,\\
    \sigma_{\rm NNLO+NNLL}^{\rm dQCD}&=& \sigma_{\rm NNLO}+ \sigma_{\rm NNLL}^{\rm dQCD} - \left.\sigma_{\rm NNLL}^{\rm dQCD}\right|_{\alphas^2}\,,
\end{eqnarray}
where $\left.\sigma_{\rm NNLL}^{\rm SCET}\right|_{\alphas^2}$ ($\left.\sigma_{\rm NNLL}^{\rm dQCD}\right|_{\alphas^2}$) is
the expansion of $ \sigma_{\rm NNLL}^{\rm SCET}$ ($ \sigma_{\rm NNLL}^{\rm dQCD}$) up to
relative order $\alphas^2$.\\
The two matched predictions are combined by simply taking their arithmetic average,
\begin{equation}
    \sigma_{\rm NNLO+NNLL}=\frac{\sigma_{\rm NNLO+NNLL}^{\rm SCET}+ \sigma_{\rm NNLO+NNLL}^{\rm dQCD}}{2}\,.
    \label{eq:sigmaavg}
\end{equation}
Finally, the addition of $_{\rm +EW}$ in the subscript corresponds to including the subleading LO and NLO contributions, i.e. to
the combination of the matched prediction with the complete-NLO corrections. So, for example:
\begin{eqnarray}
    \sigma_{\rm NNLO + EW} &&
    = \sigma_{\rm NNLO}
    + \sum_{i = 2}^3 \sigma_{\rm LO, i}
    + \sum_{j = 2}^4 \sigma_{\rm NLO, j} \, , \\
    \sigma_{\rm NNLO + NNLL + EW} &&
    = \sigma_{\rm NNLO + NNLL}
    + \sum_{i = 2}^3 \sigma_{\rm LO, i}
    + \sum_{j = 2}^4 \sigma_{\rm NLO, j} \, ,
\end{eqnarray}
and so on.

\section{Numerical results}
\label{sec:results}
In this section, we provide the state-of-the-art predictions for $t \bar t H$.
In particular, the relevant input parameters are listed
in section \ref{sec:results-input}, while
the impact of the various contributions to the $t \bar t H$ cross
section is discussed in section \ref{sec:results-contr}.
In section \ref{sec:results-errors}
we estimate different sources of theoretical errors.
\subsection{Input parameters}
\label{sec:results-input}
The input parameters for the theoretical predictions follow the recommendations of the LHC Higgs Working Group.~\footnote{See
    \url{https://twiki.cern.ch/twiki/bin/view/LHCPhysics/LHCHWG136TeVxsec}.}
In particular, we work in the five-flavour scheme, where the top quark is the only massive fermion. 
The top-quark, $Z$-, and $W$-boson masses are set to
\begin{equation}
    m_t = 172.5 \,\gev\,, \qquad
    m_W = 80.379\,\gev\,, \qquad
    m_Z = 91.1876\,\gev\,.
\end{equation}
Vector boson masses as well as the top quark mass and Yukawa coupling are renormalised in the on-shell scheme.
All particles are considered stable, and their widths are therefore neglected.\\
The value of the Fermi constant,
\begin{equation}
G_\mu = 1.16637 \times 10^{-5}\,\gev^{-2}\,,
\end{equation}
fixes the EW input scheme.
The Higgs boson mass is varied in the set of values
\begin{equation}
    m_H \in \{ 124.6, 125, 125.09, 125.38, 125.6, 126 \} \, \gev\,,
\end{equation}
while three scenarios are considered for the centre-of-mass energy $\sqrt S$:
\begin{equation}
    \sqrt S \in \{ 13, 13.6, 14\} \, \tev\,.
\end{equation}
The PDF4LHC21 parton-distribution functions (PDFs) are
employed~\cite{PDF4LHCWorkingGroup:2022cjn} for all coloured partons. Specifically, we employed the
{\tt PDF4LHC21\_40\_pdfas} set  which makes it possible to estimate the PDF-related uncertainties (using the Hessian method) together with 
those associated with $\alphas$.
Regarding the photon density, which is relevant for the EW corrections to $t \bar t H$,
a specific
choice needs to be made, as it is not included in the PDF4LHC21 combination. In particular,
the photon density prediction, based on the {\sc LuxQED}
method~\cite{Manohar:2016nzj, Manohar:2017eqh}, applied on top of the
PDF4LHC15 combination~\cite{Butterworth:2015oua} is employed.~\footnote{%
    The possibility to apply two different sets of PDFs
    is achieved through {\sc PineAppl}~\cite{Carrazza:2020gss} and
    its interface to {\sc Matrix}~\cite{Grazzini:2017mhc}.%
} \\
The central value $\mu$ of the renormalisation and
factorisation scales is fixed to half the threshold energy:
\begin{equation}
    \mu= \frac{m_H}{2} + m_t\,.
    \label{eq:scalechoice}
\end{equation}
The scale-uncertainty error is obtained by varying the two scales by a factor of 2, keeping
$0.5\le\muR/\muF\le2$ (7-point variations).

\subsection{Cross-section predictions, and impact of the various contributions}
\label{sec:results-contr}

\begin{table}[ht!]
    \centering
    \begin{tabular}{ccccccc}
        $\sqrt S$  &  $m_H$  &  $\sigma_{\rm NLO}$  &
     $\sigma_{\rm NNLO}$  &  
     $\sigma_{\rm NNLO+EW}$  &  
     $\delta_{\rm NNLO}$  &   
     $\delta_{\rm NNLO+EW}$  \\ 
$[\tev]$  &  $[\gev]$  &  $[\fb]^{+[\%]}_{-[\%]}$  &  $[\fb]^{+[\%]}_{-[\%]}$  &  $[\fb]^{+[\%]}_{-[\%]}$  &  $[\%]$  &  $[\%]$   \\
 \hline
\hline
13.0  &  124.60  &
       $ 501.7 ^{+5.8}_{-9.1}$ & 
       $ 522.8 ^{+0.9}_{-3.1}$ & 
       $ 533.8 ^{+1.1}_{-3.2}$ & 
   4.2  &
   6.4  \\
13.0  &  125.00  &
       $ 497.2 ^{+5.8}_{-9.1}$ & 
       $ 519.4 ^{+1.0}_{-3.2}$ & 
       $ 530.2 ^{+1.2}_{-3.3}$ & 
   4.5  &
   6.6  \\
13.0  &  125.09  &
       $ 496.2 ^{+5.8}_{-9.1}$ & 
       $ 517.6 ^{+0.9}_{-3.1}$ & 
       $ 528.4 ^{+1.1}_{-3.2}$ & 
   4.3  &
   6.5  \\
13.0  &  125.38  &
       $ 493.0 ^{+5.8}_{-9.1}$ & 
       $ 513.8 ^{+0.9}_{-3.1}$ & 
       $ 524.5 ^{+1.1}_{-3.2}$ & 
   4.2  &
   6.4  \\
13.0  &  125.60  &
       $ 490.5 ^{+5.8}_{-9.1}$ & 
       $ 511.0 ^{+0.9}_{-3.1}$ & 
       $ 521.7 ^{+1.1}_{-3.2}$ & 
   4.2  &
   6.4  \\
13.0  &  126.00  &
       $ 486.1 ^{+5.8}_{-9.1}$ & 
       $ 506.7 ^{+0.9}_{-3.1}$ & 
       $ 517.2 ^{+1.1}_{-3.2}$ & 
   4.2  &
   6.4  \\
\hline
13.6  &  124.60  &
       $ 563.7 ^{+5.9}_{-9.1}$ & 
       $ 586.7 ^{+0.8}_{-3.0}$ & 
       $ 598.9 ^{+1.0}_{-3.1}$ & 
   4.1  &
   6.2  \\
13.6  &  125.00  &
       $ 558.6 ^{+5.9}_{-9.1}$ & 
       $ 580.1 ^{+0.7}_{-2.9}$ & 
       $ 592.0 ^{+0.9}_{-3.1}$ & 
   3.8  &
   6.0  \\
13.6  &  125.09  &
       $ 557.5 ^{+5.9}_{-9.1}$ & 
       $ 579.7 ^{+0.8}_{-3.0}$ & 
       $ 591.7 ^{+1.0}_{-3.1}$ & 
   4.0  &
   6.1  \\
13.6  &  125.38  &
       $ 553.9 ^{+5.9}_{-9.1}$ & 
       $ 576.5 ^{+0.8}_{-3.0}$ & 
       $ 588.4 ^{+1.0}_{-3.1}$ & 
   4.1  &
   6.2  \\
13.6  &  125.60  &
       $ 551.1 ^{+5.9}_{-9.1}$ & 
       $ 573.9 ^{+0.9}_{-3.0}$ & 
       $ 585.6 ^{+1.0}_{-3.2}$ & 
   4.1  &
   6.3  \\
13.6  &  126.00  &
       $ 546.2 ^{+5.9}_{-9.1}$ & 
       $ 568.5 ^{+0.9}_{-3.0}$ & 
       $ 580.1 ^{+1.1}_{-3.2}$ & 
   4.1  &
   6.2  \\
\hline
14.0  &  124.60  &
       $ 607.0 ^{+6.0}_{-9.1}$ & 
       $ 629.1 ^{+0.6}_{-2.9}$ & 
       $ 642.1 ^{+0.8}_{-3.0}$ & 
   3.6  &
   5.8  \\
14.0  &  125.00  &
       $ 601.6 ^{+6.0}_{-9.1}$ & 
       $ 625.6 ^{+0.8}_{-3.0}$ & 
       $ 638.4 ^{+0.9}_{-3.1}$ & 
   4.0  &
   6.1  \\
14.0  &  125.09  &
       $ 600.4 ^{+6.0}_{-9.1}$ & 
       $ 622.9 ^{+0.7}_{-2.9}$ & 
       $ 635.6 ^{+0.9}_{-3.0}$ & 
   3.7  &
   5.9  \\
14.0  &  125.38  &
       $ 596.5 ^{+6.0}_{-9.1}$ & 
       $ 621.1 ^{+0.8}_{-3.0}$ & 
       $ 634.6 ^{+1.0}_{-3.2}$ & 
   4.1  &
   6.4  \\
14.0  &  125.60  &
       $ 593.6 ^{+6.0}_{-9.1}$ & 
       $ 617.7 ^{+0.8}_{-3.0}$ & 
       $ 630.2 ^{+1.0}_{-3.1}$ & 
   4.1  &
   6.2  \\
14.0  &  126.00  &
       $ 588.3 ^{+6.0}_{-9.1}$ & 
       $ 611.2 ^{+0.7}_{-3.0}$ & 
       $ 623.6 ^{+0.9}_{-3.1}$ & 
   3.9  &
   6.0  \\

    \end{tabular}
    \caption{\label{tab:tth_fo} Predictions for the process $t \bar t H$: contributions entering the fixed-order cross section. The quoted uncertainties
are obtained via 7-point scale variations.}
\end{table}
Before presenting the final results,
it is worth considering the different contributions that enter the cross section.
Starting from
those contributions that are included in fixed-order perturbation theory,
we identify in table \ref{tab:tth_fo} the impact of NNLO predictions
and of EW corrections. Specifically, we define
\begin{eqnarray}
    \delta_{\rm NNLO} = \frac{\sigma_{\rm NNLO}}{\sigma_{\rm NLO}} - 1\,,\\
    \delta_{\rm NNLO+EW} = \frac{\sigma_{\rm NNLO+EW}}{\sigma_{\rm NLO}} - 1\,,
\end{eqnarray}
i.e. the impact, relative to the NLO QCD predictions,
of the NNLO corrections alone or combined with the complete-NLO corrections.
From the table, we observe that the impact of
NNLO and EW corrections is roughly independent of the Higgs mass
and collider energy, and that both quantities amount to a few percent ($\sim 4\%$ for the NNLO QCD,
2\% for the EW). Furthermore, the inclusion of NNLO
corrections significantly reduces the theoretical uncertainties
estimated from scale variations compared to NLO,
shrinking them by roughly a factor of 3, down to $\sim 3 \%$ when the largest variation is considered.
As expected, given their small size, EW corrections have a marginal
effect on the scale-variation band.

\begin{table}
    \centering
    \begin{tabular}{ccccc|c}
        $\sqrt S$  &  $m_H$  &  $\sigma_{\rm NNLO+EW}$  &
     $\sigma_{\rm NNLO+NNLL+EW}^{\rm dQCD}$  &  
     $\sigma_{\rm NNLO+NNLL+EW}^{\rm SCET}$  &  
     $\sigma_{\rm NNLO+NNLL+EW}$  \\  
$[\tev]$  &  $[\gev]$  &  $[\fb]^{+[\%]}_{-[\%]}$  &  $[\fb]^{+[\%]}_{-[\%]}$  &  $[\fb]^{+[\%]}_{-[\%]}$  &  $[\fb]^{+[\%]}_{-[\%]}$  \\
\hline
\hline
13.0  &  124.60  &
       $ 533.8 ^{+1.1}_{-3.2}$ & 
       $ 534.4 ^{+0.6}_{-2.1}$ & 
       $ 534.1 ^{+0.5}_{-2.2}$ 
       $^{+1.6}_{-1.7}$ & 
       $ 534.2 ^{+1.6}_{-2.2}$ \\ 
13.0  &  125.00  &
       $ 530.2 ^{+1.2}_{-3.3}$ & 
       $ 530.8 ^{+0.6}_{-2.1}$ & 
       $ 530.5 ^{+0.7}_{-2.2}$ 
       $^{+1.7}_{-1.7}$ & 
       $ 530.6 ^{+1.7}_{-2.3}$ \\ 
13.0  &  125.09  &
       $ 528.4 ^{+1.1}_{-3.2}$ & 
       $ 529.0 ^{+0.6}_{-2.1}$ & 
       $ 528.7 ^{+0.5}_{-2.2}$ 
       $^{+1.6}_{-1.7}$ & 
       $ 528.8 ^{+1.6}_{-2.2}$ \\ 
13.0  &  125.38  &
       $ 524.5 ^{+1.1}_{-3.2}$ & 
       $ 525.1 ^{+0.6}_{-2.1}$ & 
       $ 524.8 ^{+0.6}_{-2.2}$ 
       $^{+1.6}_{-1.7}$ & 
       $ 524.9 ^{+1.6}_{-2.2}$ \\ 
13.0  &  125.60  &
       $ 521.7 ^{+1.1}_{-3.2}$ & 
       $ 522.2 ^{+0.6}_{-2.1}$ & 
       $ 521.9 ^{+0.5}_{-2.2}$ 
       $^{+1.6}_{-1.6}$ & 
       $ 522.1 ^{+1.6}_{-2.2}$ \\ 
13.0  &  126.00  &
       $ 517.2 ^{+1.1}_{-3.2}$ & 
       $ 517.7 ^{+0.6}_{-2.1}$ & 
       $ 517.5 ^{+0.5}_{-2.2}$ 
       $^{+1.6}_{-1.7}$ & 
       $ 517.6 ^{+1.6}_{-2.2}$ \\ 
\hline
13.6  &  124.60  &
       $ 598.9 ^{+1.0}_{-3.1}$ & 
       $ 599.5 ^{+0.5}_{-2.0}$ & 
       $ 599.2 ^{+0.3}_{-2.2}$ 
       $^{+1.5}_{-1.6}$ & 
       $ 599.3 ^{+1.5}_{-2.2}$ \\ 
13.6  &  125.00  &
       $ 592.0 ^{+0.9}_{-3.1}$ & 
       $ 592.7 ^{+0.5}_{-2.0}$ & 
       $ 592.3 ^{+0.2}_{-2.1}$ 
       $^{+1.5}_{-1.5}$ & 
       $ 592.5 ^{+1.4}_{-2.1}$ \\ 
13.6  &  125.09  &
       $ 591.7 ^{+1.0}_{-3.1}$ & 
       $ 592.3 ^{+0.5}_{-2.0}$ & 
       $ 592.0 ^{+0.2}_{-2.1}$ 
       $^{+1.5}_{-1.6}$ & 
       $ 592.1 ^{+1.5}_{-2.2}$ \\ 
13.6  &  125.38  &
       $ 588.4 ^{+1.0}_{-3.1}$ & 
       $ 589.0 ^{+0.5}_{-2.0}$ & 
       $ 588.6 ^{+0.3}_{-2.2}$ 
       $^{+1.6}_{-1.6}$ & 
       $ 588.8 ^{+1.5}_{-2.2}$ \\ 
13.6  &  125.60  &
       $ 585.6 ^{+1.0}_{-3.2}$ & 
       $ 586.2 ^{+0.5}_{-2.1}$ & 
       $ 585.9 ^{+0.4}_{-2.2}$ 
       $^{+1.6}_{-1.6}$ & 
       $ 586.0 ^{+1.6}_{-2.2}$ \\ 
13.6  &  126.00  &
       $ 580.1 ^{+1.1}_{-3.2}$ & 
       $ 580.7 ^{+0.6}_{-2.1}$ & 
       $ 580.4 ^{+0.4}_{-2.2}$ 
       $^{+1.6}_{-1.7}$ & 
       $ 580.5 ^{+1.6}_{-2.2}$ \\ 
\hline
14.0  &  124.60  &
       $ 642.1 ^{+0.8}_{-3.0}$ & 
       $ 642.7 ^{+0.5}_{-1.9}$ & 
       $ 642.4 ^{+0.2}_{-2.1}$ 
       $^{+1.3}_{-1.5}$ & 
       $ 642.6 ^{+1.3}_{-2.1}$ \\ 
14.0  &  125.00  &
       $ 638.4 ^{+0.9}_{-3.1}$ & 
       $ 639.0 ^{+0.5}_{-2.0}$ & 
       $ 638.6 ^{+0.3}_{-2.1}$ 
       $^{+1.5}_{-1.6}$ & 
       $ 638.8 ^{+1.5}_{-2.2}$ \\ 
14.0  &  125.09  &
       $ 635.6 ^{+0.9}_{-3.0}$ & 
       $ 636.3 ^{+0.5}_{-2.0}$ & 
       $ 635.9 ^{+0.2}_{-2.1}$ 
       $^{+1.4}_{-1.6}$ & 
       $ 636.1 ^{+1.4}_{-2.1}$ \\ 
14.0  &  125.38  &
       $ 634.6 ^{+1.0}_{-3.2}$ & 
       $ 635.3 ^{+0.5}_{-2.1}$ & 
       $ 634.9 ^{+0.3}_{-2.2}$ 
       $^{+1.6}_{-1.7}$ & 
       $ 635.1 ^{+1.6}_{-2.2}$ \\ 
14.0  &  125.60  &
       $ 630.2 ^{+1.0}_{-3.1}$ & 
       $ 630.9 ^{+0.5}_{-2.1}$ & 
       $ 630.4 ^{+0.3}_{-2.2}$ 
       $^{+1.6}_{-1.6}$ & 
       $ 630.7 ^{+1.5}_{-2.2}$ \\ 
14.0  &  126.00  &
       $ 623.6 ^{+0.9}_{-3.1}$ & 
       $ 624.2 ^{+0.5}_{-2.0}$ & 
       $ 623.9 ^{+0.3}_{-2.2}$ 
       $^{+1.5}_{-1.6}$ & 
       $ 624.1 ^{+1.5}_{-2.2}$ \\ 

    \end{tabular}
    \caption{%
        \label{tab:tth_fo_resum} Predictions for the process $t \bar t H$.
        The quoted uncertainties
        are obtained via 7-point scale variations. For the
        SCET predictions, two such bands are quoted,
        respectively with $\mu_h = \muR$ and $\mu_h = \muF$.
    }
\end{table}

The effects of resummation can be seen by examining table \ref{tab:tth_fo_resum}. 
We note that, regardless
of the framework, resummation changes the central prediction by
only one per mille or below, compared to NNLO, for the choice of the scale in Eq.~(\ref{eq:scalechoice}). However, its effect on the scale dependence
is substantial: the inclusion of NNLL resummation in the prediction further
reduces the scale uncertainty from 3\% at NNLO down to the level of 1.5-2\%.\footnote{ Note that the  
improved stability of the NNLO+NNLL results under scale variations is more
apparent when considered across a wider range of (parametrically) different 
scales, as in figure \ref{fig:tth_comparisons}.    }

We also see that the dQCD and
the SCET prediction\footnote{We remind the reader that, as discussed in Sec.~\ref{sec:theo-nnllb}, the soft scale is fixed to $\mu_s= Q/\bar N$ in the SCET approach.} computed with $\mu_h = \muR$ display a very similar,
but asymmetric, scale dependence, whereas
for the SCET prediction with $\mu_h = \muF$
scale variations are more symmetric, although the
overall uncertainty band stays roughly the same.
 In order to combine the two resummed calculations into 
a single result, we average the central values as in Eq.~\eqref{eq:sigmaavg} and take as the 
uncertainty band the envelope of scale variations across the two methods.
The combined prediction obtained in this way, 
which we denote as $\sigma_{\rm NNLO + NNLL + EW}$, is shown in the last column of the table.
 While the combined resummed calculation necessarily shows a larger 
uncertainty band than in either dQCD or SCET alone, the width of 
the band remains smaller than in pure fixed order. The results 
$\sigma_{\rm NNLO + NNLL + EW}$ correspond to the best prediction for the $t\bar t H$ production 
process that can be obtained to date.  


\begin{figure}[ht!]
    \includegraphics[width=0.49\textwidth]{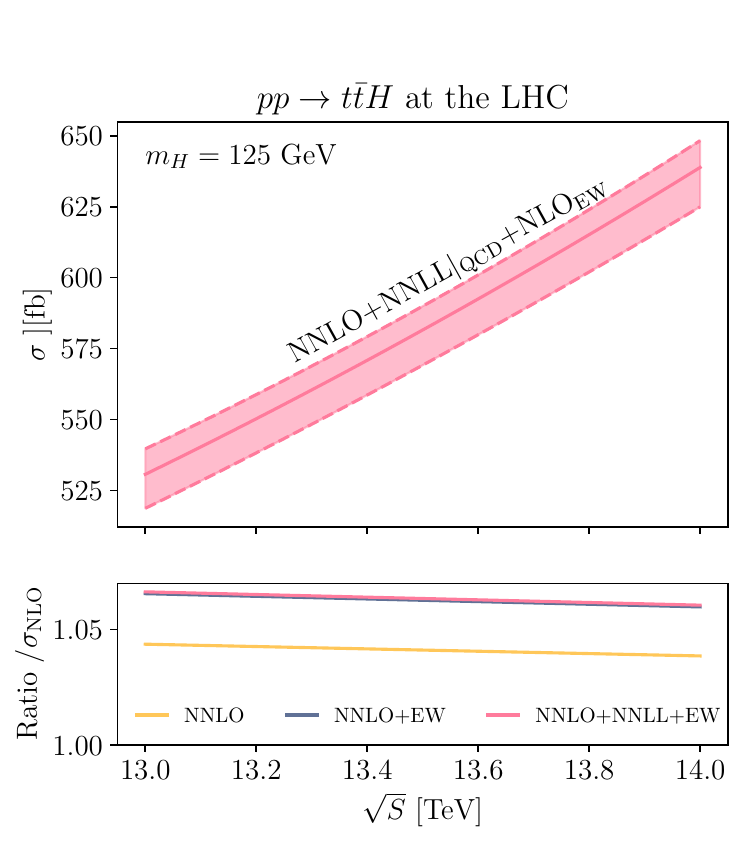}
    \includegraphics[width=0.49\textwidth]{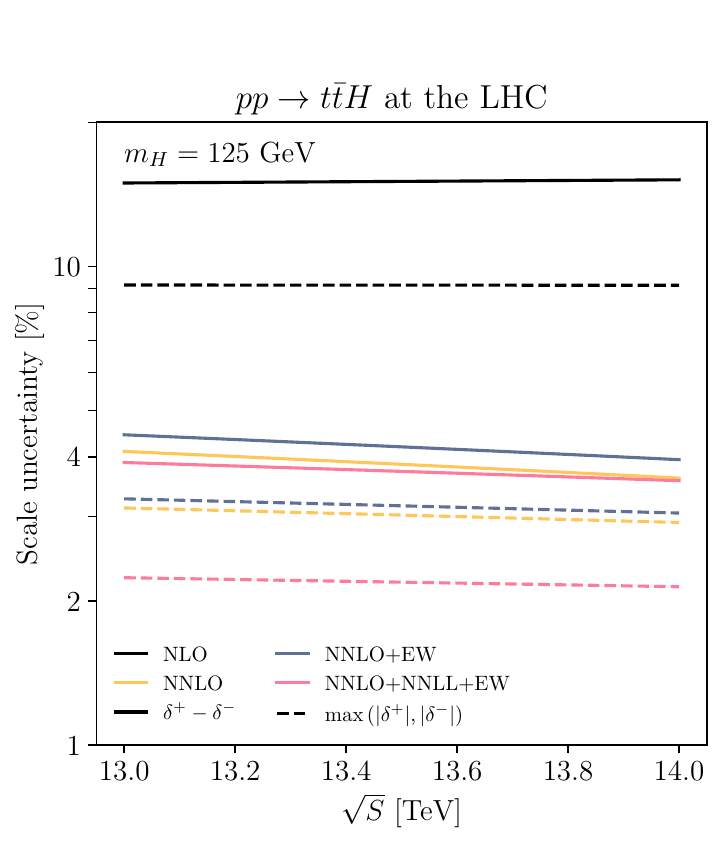}
    \caption{%
        \label{fig:xsecerr}{%
            Left: the total cross section for $t \bar t H$,
            $\sigma_{\rm NNLO + NNLL + EW}$,
            plotted as a function of the
            collider energy $\sqrt S$ for $m_H = 125 \, \gev$. The inset shows
            the relative impact of the different
            contributions with respect to $\sigma_{\rm NLO}$.
            Right:
            scale uncertainties for the cross section, computed at different
            accuracies.
            Solid lines display the total width of the scale-uncertainty band,
            while dashed lines the maximum
            variation with respect to the central prediction.%
        }%
    }
\end{figure}
The two panels of figure \ref{fig:xsecerr} plot the total cross section, the impact of the different higher-order contributions and the residual scale uncertainties 
as a function of the collider energy, for the Higgs mass value $m_H=125\,\gev$. They give a visual summary of the discussion carried out so far: the left panel shows the absolute cross section and the impact of the different 
contributions, while the right panel shows the size of theoretical uncertainties. In this case, solid lines represent the total width of the scale-uncertainty
band, while dashed lines stand for the maximum between the upper and lower scale variation.

In the following section, we will discuss the various sources of theoretical errors affecting these numbers, on top of the already mentioned scale variations.

\subsection{Residual theoretical errors}
\label{sec:results-errors}
In order to provide reliable predictions for the $t\bar t H$ cross section, a thorough estimate of all sources of theoretical errors is mandatory. In 
the previous section, we have already discussed the impact of missing higher orders in QCD, estimated via scale variations. Their smallness renders the assessment of the other possible sources of theoretical uncertainties even more crucial. We list and quantify relevant sources of theoretical uncertainties in the following. All quoted uncertainties have a negligible dependence on the specific Higgs mass and collider energies (when
varied across the range of values considered in this work). 
\begin{itemize}
    \item {\it PDF and $\alpha_s$ uncertainties:}
        uncertainties due to partonic distributions
        and the value of the strong coupling are estimated following the
        PDF4LHC prescription.
        PDF uncertainties reflect the quality
        and consistency of the data employed for the fit. They amount to
        \begin{equation}
            \Delta_{\textrm{PDF}} = 2.2\%\,.
        \end{equation}
        For what concerns the photon density,
        its minor impact together with the very precise determination
        stemming from the {\sc LuxQED}
        method renders its uncertainty negligible.
        As far as $\alphas$ is concerned, again following the PDF4LHC recommendation,
        we quote uncertainties obtained from varying
        $\alphas(m_Z)$ by an amount of  0.001 with
        respect to the default value $\alphas(m_Z) = 0.118$. This variation
        is performed both in the PDFs and in the short-distance cross section. The 
        uncertainty on the total cross section is 
        \begin{equation}
            \Delta_{\alphas} = 1.7\% \frac{\delta\alpha_s}{0.001}\,.
        \end{equation}
    \item {%
        \it Errors due to the approximation of the double-virtual contribution:%
    }
        as discussed in section \ref{sec:theo-nnlo},
        in the NNLO calculation the two-loop amplitudes are
        estimated via two approximations that are ultimately
        combined through a weighted average.
        A corresponding systematic error is assigned by means of a
        conservative procedure that takes into
        account several sources of ambiguities,
        as detailed in Ref.~\cite{Devoto:2024nhl}.
    The final error on the NNLO cross section turns out to amount to
    \begin{equation}
        \Delta_{\rm virt} = 0.9\%\,,
    \end{equation}
         and it is widely independent of
        the parameters in the range of the scan over collider energies $\sqrt{S}$ and Higgs boson masses $m_H$. 
    \item {\it Numerical uncertainties:}
      the fixed-order result has been obtained within the $q_T$-subtraction formalism. In practice, the computation is performed \cite{Grazzini:2017mhc} by introducing a technical
      cut-off \mbox{$r_{\mathrm{cut}} = q_T^{\mathrm{cut}}/Q$} on the dimensionless variable $q_T/Q$,
      where $q_T\, (Q)$ is the transverse momentum (invariant mass) of the $t \bar t H$ system.
      The final result, which corresponds to the limit \mbox{$r_{\mathrm{cut}} \to 0$}, is extracted by simultaneously
      computing the cross section at fixed values of $r_{\mathrm{cut}}$ and then performing an extrapolation to that limit. 
      The error associated with this procedure, which combines statistical and extrapolation uncertainties,
      varies slightly between the setups, but is always small ($\mathcal{O}(0.3\%)$) compared to
      the other sources of theoretical uncertainties. 
    \item {\it Ambiguities in the resummation approach:}
        as we have discussed in section \ref{sec:results-contr},
        in particular when discussing table \ref{tab:tth_fo_resum}
        the two different resummation procedures employed have a
        marginal effect on the total cross section,
        while they reduce the scale uncertainty band.
        Specifically,
        the effect on the cross section is at most at the 0.1\% level.
        Moreover,
        the scale uncertainty band takes into account all scale variations
        from the two methods, and has thus to be regarded as very conservative
    \item {\it Uncertainties related to the top-quark mass value}:
        we estimate the dependence of the cross section on the
        top-quark mass by reporting
        how the cross section varies when
        $m_t$ is changed by $1 \, \gev$
        with respect to the reference value.
        The top-quark mass enters both the kinematics part of the cross section (the dominant impact is from the phase space) 
        and the top-quark Yukawa coupling, and the
        two effects have opposite sign. Remarkably, they tend to cancel almost exactly at the energies we considered, so the
        top-mass dependence can be neglected for the SM predictions, when the relation $y_t = \frac{\sqrt 2 m_t}{v}$ is enforced.\\
        Still, it is also worth considering the case when the top-quark mass and the top-quark Yukawa coupling are varied independently. In this case, we obtain
        \begin{equation}
            \Delta_{y_t} = 1.1\%\frac{v}{\sqrt 2} \frac{\delta y_t}{1 \gev}\,, \qquad \Delta_{m_t} = -1.1\% \frac{\delta m_t}{1\gev} \,.
        \end{equation}
        The opposite sign of $\Delta_{y_t}$ and $\Delta_{m_t}$ reflects the opposite slope of the cross section when $y_t$ or $m_t$ are increased.\\
        We note that, strictly speaking, this procedure is inconsistent when
             EW corrections are included. The effect of $y_t$ and $m_t$ variations, therefore, is assessed by neglecting them.
            However, given their rather small impact,
            EW corrections do not alter significantly
            the dependence of the cross section on $y_t$.
    \item {\it Uncertainties related to the top-quark mass renormalisation scheme}:
        As mentioned in Sec.~\ref{sec:results-input}, we worked with the top-quark mass and Yukawa couplings renormalised in the on-shell scheme. An alternative scheme to employ is
        $\overline{\textrm{MS}}$, which is usually the reference scheme for the case of lighter heavy quarks (e.g. the bottom), since it resums to all orders
        logarithms involving 
        the ratio $\frac{m^2}{\mu_R^2}$, $m$ being the heavy-quark mass (see e.g. Refs.~\cite{Dittmaier:2003ej,Wiesemann:2014ioa,Pagani:2020rsg}). In
        the case of $t\bar tH$, due to the large top-quark mass, these effects are expected to be negligible. Furthermore,
        any effect related to the renormalisation scheme must be higher-order with respect to the perturbative order at which predictions are computed.
        Results presented in Ref.~\cite{Saibel:2021krs}, where the $t\bar t H$ cross section is computed at NLO in both schemes,
        help us giving a more quantitative statement. If we consider the total rate at NLO, changing the scheme from on-shell to $\overline{\textrm{MS}}$
        amounts to a 1\% effect when in the latter scheme the top mass is evaluated at the (fixed) scale $m_t$. At NNLO such an effect is expected
        to be further reduced, hence negligible. We stress, however, that if differential observables are considered, larger effects may appear.
    \item {\it Uncertainties due to missing higher-order EW corrections:}
        since in the $G_\mu$ scheme the EW coupling is kept fixed,
        the relative effect of scale variations for
        EW corrections is identical to the LO contribution,
        and it does not cover missing higher orders in $\alpha$.
        Here we provide an argument to estimate NNLO$_2$, the first contribution where EW effects enter at NNLO, and which corresponds to 
        $O(\alpha \alphas)$ corrections to $\sigma_{\rm LO}$.
        In order to have a rough estimate of possible effects at this order,
          the typical procedure is to study the difference between additive and multiplicative
                combinations of NLO QCD and NLO EW corrections. In a multiplicative combination, an extra term
                of $O(\alpha \alphas)$ appears, which improves the scale dependence of NLO EW corrections. In our case, we consider
                the NLO QCD and NLO EW $K$ factors, whose numerical value is the following:
                \begin{equation}
                    K_{\rm NLO QCD} \equiv \frac{\sigma_{\rm NLO}}{\sigma_{\rm LO}}=1.26\,,\qquad
                    K_{\rm NLO EW} \equiv \frac{\sigma_{\rm LO+EW}}{\sigma_{\rm LO}}=1.02\,.
                \end{equation}
                The additive and multiplicative combinations are defined by
                \begin{eqnarray}
                    K_{\rm NLO QCD+EW} &\equiv& K_{\rm NLO QCD} + K_{\rm NLO EW} -1 = 1.28\,,\\
                    K_{\rm NLO QCD\times EW} &\equiv& K_{\rm NLO QCD} \times K_{\rm NLO EW}=K_{\rm NLO QCD+EW} + 0.005 \,.\nonumber
                \end{eqnarray}
                Where the extra $0.005$ (0.5\%) precisely corresponds to the extra $O(\alpha \alphas)$ term. This term gives a rough estimate of the NNLO$_2$ contribution.
        Considering that the overall impact of such an uncertainty is further diluted by the large NLO QCD corrections,
        we can conclude that missing higher-order EW contributions will amount
        at most to few per mille of the final prediction, and therefore can be considered as a subleading source of uncertainty.
\end{itemize}
To clarify the previous discussion and show its application to a practical case, we report our state-of-the-art prediction, equipped with the dominant sources of uncertainties, for the $t\bar t H$ total cross section in the SM at $\sqrt S= 13.6 \tev$ and $m_H= 125.09 \gev$: 
\begin{equation}
    \sigma_{\rm NNLO+NNLL+EW}^{\sqrt S= 13.6 \tev,\, m_H= 125.09 \gev} = 592.1\,\fb\; \underbrace{ ^{+1.5\%}_{-2.2\%}}_{\Delta_\mu}\;
    \underbrace{\pm 2.2\%}_{\Delta_{\rm PDF}} \;
    \underbrace{\pm 1.7\%}_{\Delta_{\alphas}} \;
    \underbrace{\pm 0.9\%}_{\Delta_{\rm virt}}\,,
\end{equation}
where we \emph{assumed} the parametric uncertainty on $\alphas$, $\delta \alpha_s=0.001$.

\section{Conclusions}
\label{sec:concl}

The $t \bar t H$ production process is a sensitive probe of the top-quark Yukawa coupling
and its cross section represents a key observable in LHC physics. 
In this paper, we have presented a state-of-the-art
computation of this observable within the SM. This computation combines the recently obtained NNLO QCD corrections with
NNLL soft-gluon resummation and complete-NLO corrections into a full NNLO+NNLL+EW
prediction.

The starting point of our calculation is the recent NNLO QCD computation of Ref.~\cite{Devoto:2024nhl}. The missing two-loop amplitudes
 are derived therein by using two independent approximations that are ultimately combined to obtain an estimate of the finite part of the two-loop virtual contribution and its uncertainty. All the remaining ingredients of the calculation are evaluated exactly.
The ensuing NNLO QCD result is combined with soft-gluon resummation up to NNLL accuracy.

From the perspective of perturbative QCD, a particularly interesting outcome of our work is the first-ever comparison
of SCET and dQCD-based soft-gluon resummations at NNLO+NNLL order, for a process involving the non-trivial colour structure characteristic of four coloured partons in the Born level amplitude. Although the two methods share a common starting point, namely the factorisation of the partonic cross section in the soft gluon emission limit, the implementation of renormalisation-group improved perturbation theory underlying the resummations differs, so that the SCET formulas contain some corrections that are considered N$^3$LL and higher in dQCD and vice  versa. In spite of these systematic differences between the two frameworks, the numerical results agree remarkably well at NNLO+NNLL in QCD,  as clearly seen in the left-hand panel of figure~\ref{fig:tth_comparisons}. In both
cases the resummation stabilizes scale uncertainties compared to NNLO alone, especially when considered across a wide range of (parametrically) different scales, as is apparent from the right-hand panel of the same figure. We have thus taken a conservative approach to residual resummation
errors, taking into account the systematic differences between dQCD and SCET in addition to scale variations.

The NNLO+NNLL QCD results are eventually combined with the complete-NLO corrections, whose effect, although small, must be included at this level of precision.
From a purely phenomenological perspective, our main results can be
found in table \ref{tab:tth_fo_resum}, which shows NNLO+NNLL+EW predictions as a function of the LHC collider energy and Higgs mass, including an
estimate of uncertainties from even higher-order QCD corrections. Remarkably, these
uncertainty estimates are at the $\pm1$-2 percent level. Other sources of theoretical uncertainty have been discussed and  quantified in section \ref{sec:results-errors} -- those related to PDFs and $\alpha_s$ are currently the dominant ones, followed by those stemming from the approximation of two loop amplitudes.

\section{Citation policy}
\label{sec:cite}
The present work consists of contributions from different collaborations, whose work needs to be acknowledged. If the results are employed for scientific
publications, together with this work, one should cite 
Refs.~\cite{Frixione:2014qaa,Zhang:2014gcy,Frixione:2015zaa,Kulesza:2015vda,Broggio:2015lya,Broggio:2016lfj,Kulesza:2017ukk,Ju:2019lwp,Kulesza:2018tqz,Broggio:2019ewu,Kulesza:2020nfh,Catani:2022mfv,Devoto:2024nhl}.
The relevant {\sc BibTex} keys are:
{\small
\begin{verbatim}
Frixione:2014qaa,Frixione:2015zaa,Kulesza:2015vda,Broggio:2015lya,Broggio:2016lfj,
Kulesza:2017ukk,Frederix:2018nkq,Kulesza:2018tqz,Broggio:2019ewu,Kulesza:2020nfh,
Catani:2022mfv,Devoto:2024nhl
\end{verbatim}
}
\section*{Acknowledgements}
This work has been carried out within the LHC Higgs Working Group, as a contribution to the Report 5 collection.
We thank the members of the working group for various
discussions on this topic, and for pushing us to pursue the work
presented here. In particular, MW and MZ are specially grateful 
to Judit Katzy, Davide Valsecchi, Josh McFayden and Sergio Sanchez Cruz for their work as experimental conveners
within the $t\bar t H$ subgroup.\\

The work of S.D. has been funded by the European Union (ERC,
MultiScaleAmp, Grant Agreement No. 101078449). Views and opinions
expressed are however those of the author(s) only and do not
necessarily reflect those of the European Union or the European
Research Council Executive Agency.  Neither the European Union nor the
granting authority can be held responsible for them.
R.F. acknowledges the support by the Swedish Research Council under
contract number 2020-04423.
A.K. acknowledges the support by the German Research Foundation (DFG)
under grants KU 3103/1 and KU 3103/2.
T.S. kindly acknowledges the support of the Polish National Science Center (NCN) grant No. 2021/43/D/ST2/03375.
The work of M.W. was supported by the German Research Foundation
(DFG) under grant 396021762 - TRR 257: \textit{Particle Physics
  Phenomenology after the Higgs Discovery}.
D.P and M.Z. acknowledge financial support by the Italian Ministry of University and Research (MUR) through the PRIN2022 Grant 2022EZ3S3F, funded by the European Union – NextGenerationEU.

\addcontentsline{toc}{section}{References}
\bibliographystyle{JHEP}
\bibliography{biblio}

\providecommand{\href}[2]{#2}\begingroup\raggedright\begin{thebibliography}{100}

\bibitem{Aad:2012tfa}
{\scshape ATLAS} collaboration, G.~Aad et~al., \emph{{Observation of a new
  particle in the search for the Standard Model Higgs boson with the ATLAS
  detector at the LHC}},
  \href{http://dx.doi.org/10.1016/j.physletb.2012.08.020}{\emph{Phys. Lett. B}
  {\bf 716} (2012) 1--29}, [\href{http://arxiv.org/abs/1207.7214}{{\tt
  1207.7214}}].

\bibitem{Chatrchyan:2012ufa}
{\scshape CMS} collaboration, S.~Chatrchyan et~al., \emph{{Observation of a New
  Boson at a Mass of 125 GeV with the CMS Experiment at the LHC}},
  \href{http://dx.doi.org/10.1016/j.physletb.2012.08.021}{\emph{Phys. Lett. B}
  {\bf 716} (2012) 30--61}, [\href{http://arxiv.org/abs/1207.7235}{{\tt
  1207.7235}}].

\bibitem{ATLAS:2022vkf}
{\scshape ATLAS} collaboration, G.~Aad et~al., \emph{{A detailed map of Higgs
  boson interactions by the ATLAS experiment ten years after the discovery}},
  \href{http://dx.doi.org/10.1038/s41586-022-04893-w}{\emph{Nature} {\bf 607}
  (2022) 52--59}, [\href{http://arxiv.org/abs/2207.00092}{{\tt 2207.00092}}].

\bibitem{CMS:2022dwd}
{\scshape CMS} collaboration, A.~Tumasyan et~al., \emph{{A portrait of the
  Higgs boson by the CMS experiment ten years after the discovery.}},
  \href{http://dx.doi.org/10.1038/s41586-022-04892-x}{\emph{Nature} {\bf 607}
  (2022) 60--68}, [\href{http://arxiv.org/abs/2207.00043}{{\tt 2207.00043}}].

\bibitem{CMS:2014wdm}
{\scshape CMS} collaboration, S.~Chatrchyan et~al., \emph{{Evidence for the 125
  GeV Higgs boson decaying to a pair of $\tau$ leptons}},
  \href{http://dx.doi.org/10.1007/JHEP05(2014)104}{\emph{JHEP} {\bf 05} (2014)
  104}, [\href{http://arxiv.org/abs/1401.5041}{{\tt 1401.5041}}].

\bibitem{CMS:2017odg}
{\scshape CMS} collaboration, A.~M. Sirunyan et~al., \emph{{Evidence for the
  Higgs boson decay to a bottom quark\textendash{}antiquark pair}},
  \href{http://dx.doi.org/10.1016/j.physletb.2018.02.050}{\emph{Phys. Lett. B}
  {\bf 780} (2018) 501--532}, [\href{http://arxiv.org/abs/1709.07497}{{\tt
  1709.07497}}].

\bibitem{ATLAS:2017ztq}
{\scshape ATLAS} collaboration, M.~Aaboud et~al., \emph{{Evidence for the
  associated production of the Higgs boson and a top quark pair with the ATLAS
  detector}}, \href{http://dx.doi.org/10.1103/PhysRevD.97.072003}{\emph{Phys.
  Rev. D} {\bf 97} (2018) 072003}, [\href{http://arxiv.org/abs/1712.08891}{{\tt
  1712.08891}}].

\bibitem{CMS:2018fdh}
{\scshape CMS} collaboration, A.~M. Sirunyan et~al., \emph{{Evidence for
  associated production of a Higgs boson with a top quark pair in final states
  with electrons, muons, and hadronically decaying $\tau$ leptons at $\sqrt{s}
  =$ 13 TeV}}, \href{http://dx.doi.org/10.1007/JHEP08(2018)066}{\emph{JHEP}
  {\bf 08} (2018) 066}, [\href{http://arxiv.org/abs/1803.05485}{{\tt
  1803.05485}}].

\bibitem{CMS:2018uxb}
{\scshape CMS} collaboration, A.~M. Sirunyan et~al., \emph{{Observation of
  $t\bar{t}H$ production}},
  \href{http://dx.doi.org/10.1103/PhysRevLett.120.231801}{\emph{Phys. Rev.
  Lett.} {\bf 120} (2018) 231801}, [\href{http://arxiv.org/abs/1804.02610}{{\tt
  1804.02610}}].

\bibitem{ATLAS:2018mme}
{\scshape ATLAS} collaboration, M.~Aaboud et~al., \emph{{Observation of Higgs
  boson production in association with a top quark pair at the LHC with the
  ATLAS detector}},
  \href{http://dx.doi.org/10.1016/j.physletb.2018.07.035}{\emph{Phys. Lett. B}
  {\bf 784} (2018) 173--191}, [\href{http://arxiv.org/abs/1806.00425}{{\tt
  1806.00425}}].

\bibitem{CMS:2018nsn}
{\scshape CMS} collaboration, A.~M. Sirunyan et~al., \emph{{Observation of
  Higgs boson decay to bottom quarks}},
  \href{http://dx.doi.org/10.1103/PhysRevLett.121.121801}{\emph{Phys. Rev.
  Lett.} {\bf 121} (2018) 121801}, [\href{http://arxiv.org/abs/1808.08242}{{\tt
  1808.08242}}].

\bibitem{ATLAS:2018kot}
{\scshape ATLAS} collaboration, M.~Aaboud et~al., \emph{{Observation of $H
  \rightarrow b\bar{b}$ decays and $VH$ production with the ATLAS detector}},
  \href{http://dx.doi.org/10.1016/j.physletb.2018.09.013}{\emph{Phys. Lett. B}
  {\bf 786} (2018) 59--86}, [\href{http://arxiv.org/abs/1808.08238}{{\tt
  1808.08238}}].

\bibitem{ATLAS:2018ynr}
{\scshape ATLAS} collaboration, M.~Aaboud et~al., \emph{{Cross-section
  measurements of the Higgs boson decaying into a pair of $\tau$-leptons in
  proton-proton collisions at $\sqrt{s}=13$ TeV with the ATLAS detector}},
  \href{http://dx.doi.org/10.1103/PhysRevD.99.072001}{\emph{Phys. Rev. D} {\bf
  99} (2019) 072001}, [\href{http://arxiv.org/abs/1811.08856}{{\tt
  1811.08856}}].

\bibitem{CMS:2020xwi}
{\scshape CMS} collaboration, A.~M. Sirunyan et~al., \emph{{Evidence for Higgs
  boson decay to a pair of muons}},
  \href{http://dx.doi.org/10.1007/JHEP01(2021)148}{\emph{JHEP} {\bf 01} (2021)
  148}, [\href{http://arxiv.org/abs/2009.04363}{{\tt 2009.04363}}].

\bibitem{CMS:2022psv}
{\scshape CMS} collaboration, A.~Tumasyan et~al., \emph{{Search for Higgs Boson
  Decay to a Charm Quark-Antiquark Pair in Proton-Proton Collisions at
  s=13\,\,TeV}},
  \href{http://dx.doi.org/10.1103/PhysRevLett.131.061801}{\emph{Phys. Rev.
  Lett.} {\bf 131} (2023) 061801}, [\href{http://arxiv.org/abs/2205.05550}{{\tt
  2205.05550}}].

\bibitem{ATLAS:2024yzu}
{\scshape ATLAS} collaboration, G.~Aad et~al., \emph{{Measurements of $WH$ and
  $ZH$ production with Higgs boson decays into bottom quarks and direct
  constraints on the charm Yukawa coupling in $13\,\mathrm{TeV}$$pp$ collisions
  with the ATLAS detector}},  \href{http://arxiv.org/abs/2410.19611}{{\tt
  2410.19611}}.

\bibitem{Cao:2016wib}
Q.-H. Cao, S.-L. Chen and Y.~Liu, \emph{{Probing Higgs Width and Top Quark
  Yukawa Coupling from $t\bar{t}H$ and $t\bar{t}t\bar{t}$ Productions}},
  \href{http://dx.doi.org/10.1103/PhysRevD.95.053004}{\emph{Phys. Rev. D} {\bf
  95} (2017) 053004}, [\href{http://arxiv.org/abs/1602.01934}{{\tt
  1602.01934}}].

\bibitem{Cao:2019ygh}
Q.-H. Cao, S.-L. Chen, Y.~Liu, R.~Zhang and Y.~Zhang, \emph{{Limiting top
  quark-Higgs boson interaction and Higgs-boson width from multitop
  productions}},
  \href{http://dx.doi.org/10.1103/PhysRevD.99.113003}{\emph{Phys. Rev. D} {\bf
  99} (2019) 113003}, [\href{http://arxiv.org/abs/1901.04567}{{\tt
  1901.04567}}].

\bibitem{CMS:2019rvj}
{\scshape CMS} collaboration, A.~M. Sirunyan et~al., \emph{{Search for
  production of four top quarks in final states with same-sign or multiple
  leptons in proton-proton collisions at $\sqrt{s}=$ 13 TeV}},
  \href{http://dx.doi.org/10.1140/epjc/s10052-019-7593-7}{\emph{Eur. Phys. J.
  C} {\bf 80} (2020) 75}, [\href{http://arxiv.org/abs/1908.06463}{{\tt
  1908.06463}}].

\bibitem{Kuhn:2013zoa}
J.~H. K\"uhn, A.~Scharf and P.~Uwer, \emph{{Weak Interactions in Top-Quark Pair
  Production at Hadron Colliders: An Update}},
  \href{http://dx.doi.org/10.1103/PhysRevD.91.014020}{\emph{Phys. Rev. D} {\bf
  91} (2015) 014020}, [\href{http://arxiv.org/abs/1305.5773}{{\tt 1305.5773}}].

\bibitem{Martini:2019lsi}
T.~Martini and M.~Schulze, \emph{{Electroweak loops as a probe of new physics
  in $ t\overline{t} $ production at the LHC}},
  \href{http://dx.doi.org/10.1007/JHEP04(2020)017}{\emph{JHEP} {\bf 04} (2020)
  017}, [\href{http://arxiv.org/abs/1911.11244}{{\tt 1911.11244}}].

\bibitem{CMS:2020djy}
{\scshape CMS} collaboration, A.~M. Sirunyan et~al., \emph{{Measurement of the
  top quark Yukawa coupling from $\mathrm{t\bar{t}}$ kinematic distributions in
  the dilepton final state in proton-proton collisions at $\sqrt{s}=$ 13 TeV}},
  \href{http://dx.doi.org/10.1103/PhysRevD.102.092013}{\emph{Phys. Rev. D} {\bf
  102} (2020) 092013}, [\href{http://arxiv.org/abs/2009.07123}{{\tt
  2009.07123}}].

\bibitem{Martini:2021uey}
T.~Martini, R.-Q. Pan, M.~Schulze and M.~Xiao, \emph{{Probing the CP structure
  of the top quark Yukawa coupling: Loop sensitivity versus on-shell
  sensitivity}},
  \href{http://dx.doi.org/10.1103/PhysRevD.104.055045}{\emph{Phys. Rev. D} {\bf
  104} (2021) 055045}, [\href{http://arxiv.org/abs/2104.04277}{{\tt
  2104.04277}}].

\bibitem{Maltoni:2024wyh}
F.~Maltoni, D.~Pagani and S.~Tentori, \emph{{Top-quark pair production as a
  probe of light top-philic scalars and anomalous Higgs interactions}},
  \href{http://dx.doi.org/10.1007/JHEP09(2024)098}{\emph{JHEP} {\bf 09} (2024)
  098}, [\href{http://arxiv.org/abs/2406.06694}{{\tt 2406.06694}}].

\bibitem{Demartin:2015uha}
F.~Demartin, F.~Maltoni, K.~Mawatari and M.~Zaro, \emph{{Higgs production in
  association with a single top quark at the LHC}},
  \href{http://dx.doi.org/10.1140/epjc/s10052-015-3475-9}{\emph{Eur. Phys. J.
  C} {\bf 75} (2015) 267}, [\href{http://arxiv.org/abs/1504.00611}{{\tt
  1504.00611}}].

\bibitem{Demartin:2016axk}
F.~Demartin, B.~Maier, F.~Maltoni, K.~Mawatari and M.~Zaro, \emph{{tWH
  associated production at the LHC}},
  \href{http://dx.doi.org/10.1140/epjc/s10052-017-4601-7}{\emph{Eur. Phys. J.
  C} {\bf 77} (2017) 34}, [\href{http://arxiv.org/abs/1607.05862}{{\tt
  1607.05862}}].

\bibitem{ATLAS:2020ior}
{\scshape ATLAS} collaboration, G.~Aad et~al., \emph{{$CP$ Properties of Higgs
  Boson Interactions with Top Quarks in the $t\bar{t}H$ and $tH$ Processes
  Using $H \rightarrow \gamma\gamma$ with the ATLAS Detector}},
  \href{http://dx.doi.org/10.1103/PhysRevLett.125.061802}{\emph{Phys. Rev.
  Lett.} {\bf 125} (2020) 061802}, [\href{http://arxiv.org/abs/2004.04545}{{\tt
  2004.04545}}].

\bibitem{CMS:2020cga}
{\scshape CMS} collaboration, A.~M. Sirunyan et~al., \emph{{Measurements of
  ${t\bar{t}}H$ Production and the CP Structure of the Yukawa Interaction
  between the Higgs Boson and Top Quark in the Diphoton Decay Channel}},
  \href{http://dx.doi.org/10.1103/PhysRevLett.125.061801}{\emph{Phys. Rev.
  Lett.} {\bf 125} (2020) 061801}, [\href{http://arxiv.org/abs/2003.10866}{{\tt
  2003.10866}}].

\bibitem{CMS:2022dbt}
{\scshape CMS} collaboration, A.~Tumasyan et~al., \emph{{Search for $CP$
  violation in $t\bar{t}H$ and $tH$ production in multilepton channels in
  proton-proton collisions at $\sqrt{s}$ = 13 TeV}},
  \href{http://dx.doi.org/10.1007/JHEP07(2023)092}{\emph{JHEP} {\bf 07} (2023)
  092}, [\href{http://arxiv.org/abs/2208.02686}{{\tt 2208.02686}}].

\bibitem{ATLAS:2022tnm}
{\scshape ATLAS} collaboration, G.~Aad et~al., \emph{{Measurement of the
  properties of Higgs boson production at $\sqrt{s} = 13$ TeV in the
  $H\to\gamma\gamma$ channel using $139$ fb$^{-1}$ of $pp$ collision data with
  the ATLAS experiment}},
  \href{http://dx.doi.org/10.1007/JHEP07(2023)088}{\emph{JHEP} {\bf 07} (2023)
  088}, [\href{http://arxiv.org/abs/2207.00348}{{\tt 2207.00348}}].

\bibitem{ATLAS:2023cbt}
{\scshape ATLAS} collaboration, G.~Aad et~al., \emph{{Probing the CP nature of
  the top-Higgs Yukawa coupling in $t\bar{t}H$ and $tH$ events with $H\to bb$
  decays using the ATLAS detector at the LHC}},
  \href{http://dx.doi.org/10.1016/j.physletb.2024.138469}{\emph{Phys. Lett. B}
  {\bf 849} (2024) 138469}, [\href{http://arxiv.org/abs/2303.05974}{{\tt
  2303.05974}}].

\bibitem{CMS:2024fdo}
{\scshape CMS} collaboration, A.~Hayrapetyan et~al., \emph{{Measurement of the
  $t\bar{t}H$ and tH production rates in the $H\to b\bar{b}$ decay channel
  using proton-proton collision data at $\sqrt{s}$ = 13 TeV}},
  \href{http://arxiv.org/abs/2407.10896}{{\tt 2407.10896}}.

\bibitem{Beenakker:2001rj}
W.~Beenakker, S.~Dittmaier, M.~Kramer, B.~Plumper, M.~Spira and P.~M. Zerwas,
  \emph{{Higgs radiation off top quarks at the Tevatron and the LHC}},
  \href{http://dx.doi.org/10.1103/PhysRevLett.87.201805}{\emph{Phys. Rev.
  Lett.} {\bf 87} (2001) 201805},
  [\href{http://arxiv.org/abs/hep-ph/0107081}{{\tt hep-ph/0107081}}].

\bibitem{Reina:2001sf}
L.~Reina and S.~Dawson, \emph{{Next-to-leading order results for t anti-t h
  production at the Tevatron}},
  \href{http://dx.doi.org/10.1103/PhysRevLett.87.201804}{\emph{Phys. Rev.
  Lett.} {\bf 87} (2001) 201804},
  [\href{http://arxiv.org/abs/hep-ph/0107101}{{\tt hep-ph/0107101}}].

\bibitem{Reina:2001bc}
L.~Reina, S.~Dawson and D.~Wackeroth, \emph{{QCD corrections to associated t
  anti-t h production at the Tevatron}},
  \href{http://dx.doi.org/10.1103/PhysRevD.65.053017}{\emph{Phys. Rev. D} {\bf
  65} (2002) 053017}, [\href{http://arxiv.org/abs/hep-ph/0109066}{{\tt
  hep-ph/0109066}}].

\bibitem{Beenakker:2002nc}
W.~Beenakker, S.~Dittmaier, M.~Kramer, B.~Plumper, M.~Spira and P.~M. Zerwas,
  \emph{{NLO QCD corrections to t anti-t H production in hadron collisions}},
  \href{http://dx.doi.org/10.1016/S0550-3213(03)00044-0}{\emph{Nucl. Phys. B}
  {\bf 653} (2003) 151--203}, [\href{http://arxiv.org/abs/hep-ph/0211352}{{\tt
  hep-ph/0211352}}].

\bibitem{Dawson:2002tg}
S.~Dawson, L.~H. Orr, L.~Reina and D.~Wackeroth, \emph{{Associated top quark
  Higgs boson production at the LHC}},
  \href{http://dx.doi.org/10.1103/PhysRevD.67.071503}{\emph{Phys. Rev. D} {\bf
  67} (2003) 071503}, [\href{http://arxiv.org/abs/hep-ph/0211438}{{\tt
  hep-ph/0211438}}].

\bibitem{Frixione:2014qaa}
S.~Frixione, V.~Hirschi, D.~Pagani, H.~S. Shao and M.~Zaro, \emph{{Weak
  corrections to Higgs hadroproduction in association with a top-quark pair}},
  \href{http://dx.doi.org/10.1007/JHEP09(2014)065}{\emph{JHEP} {\bf 09} (2014)
  065}, [\href{http://arxiv.org/abs/1407.0823}{{\tt 1407.0823}}].

\bibitem{Zhang:2014gcy}
Y.~Zhang, W.-G. Ma, R.-Y. Zhang, C.~Chen and L.~Guo, \emph{{QCD NLO and EW NLO
  corrections to $t\bar{t}H$ production with top quark decays at hadron
  collider}},
  \href{http://dx.doi.org/10.1016/j.physletb.2014.09.022}{\emph{Phys. Lett. B}
  {\bf 738} (2014) 1--5}, [\href{http://arxiv.org/abs/1407.1110}{{\tt
  1407.1110}}].

\bibitem{Frixione:2015zaa}
S.~Frixione, V.~Hirschi, D.~Pagani, H.~S. Shao and M.~Zaro, \emph{{Electroweak
  and QCD corrections to top-pair hadroproduction in association with heavy
  bosons}}, \href{http://dx.doi.org/10.1007/JHEP06(2015)184}{\emph{JHEP} {\bf
  06} (2015) 184}, [\href{http://arxiv.org/abs/1504.03446}{{\tt 1504.03446}}].

\bibitem{Frederix:2018nkq}
R.~Frederix, S.~Frixione, V.~Hirschi, D.~Pagani, H.~S. Shao and M.~Zaro,
  \emph{{The automation of next-to-leading order electroweak calculations}},
  \href{http://dx.doi.org/10.1007/JHEP07(2018)185}{\emph{JHEP} {\bf 07} (2018)
  185}, [\href{http://arxiv.org/abs/1804.10017}{{\tt 1804.10017}}].

\bibitem{Catani:2021cbl}
S.~Catani, I.~Fabre, M.~Grazzini and S.~Kallweit, \emph{{${t {{\bar{t}}}H}$
  production at NNLO: the flavour off-diagonal channels}},
  \href{http://dx.doi.org/10.1140/epjc/s10052-021-09247-w}{\emph{Eur. Phys. J.
  C} {\bf 81} (2021) 491}, [\href{http://arxiv.org/abs/2102.03256}{{\tt
  2102.03256}}].

\bibitem{Catani:2022mfv}
S.~Catani, S.~Devoto, M.~Grazzini, S.~Kallweit, J.~Mazzitelli and C.~Savoini,
  \emph{{Higgs Boson Production in Association with a Top-Antitop Quark Pair in
  Next-to-Next-to-Leading Order QCD}},
  \href{http://dx.doi.org/10.1103/PhysRevLett.130.111902}{\emph{Phys. Rev.
  Lett.} {\bf 130} (2023) 111902}, [\href{http://arxiv.org/abs/2210.07846}{{\tt
  2210.07846}}].

\bibitem{Devoto:2024nhl}
S.~Devoto, M.~Grazzini, S.~Kallweit, J.~Mazzitelli and C.~Savoini,
  \emph{{Precise predictions for $t \bar t H$ production at the LHC: inclusive
  cross section and differential distributions}},
  \href{http://arxiv.org/abs/2411.15340}{{\tt 2411.15340}}.

\bibitem{FebresCordero:2023pww}
F.~Febres~Cordero, G.~Figueiredo, M.~Kraus, B.~Page and L.~Reina,
  \emph{{Two-loop master integrals for leading-color $ pp\to t\overline{t}H $
  amplitudes with a light-quark loop}},
  \href{http://dx.doi.org/10.1007/JHEP07(2024)084}{\emph{JHEP} {\bf 07} (2024)
  084}, [\href{http://arxiv.org/abs/2312.08131}{{\tt 2312.08131}}].

\bibitem{Agarwal:2024jyq}
B.~Agarwal, G.~Heinrich, S.~P. Jones, M.~Kerner, S.~Y. Klein, J.~Lang et~al.,
  \emph{{Two-loop amplitudes for $ t\overline{t}H $ production: the
  quark-initiated N$_{f}$-part}},
  \href{http://dx.doi.org/10.1007/JHEP05(2024)013}{\emph{JHEP} {\bf 05} (2024)
  013}, [\href{http://arxiv.org/abs/2402.03301}{{\tt 2402.03301}}].

\bibitem{Wang:2024pmv}
G.~Wang, T.~Xia, L.~L. Yang and X.~Ye, \emph{{Two-loop QCD amplitudes for $
  t\overline{t}H $ production from boosted limit}},
  \href{http://dx.doi.org/10.1007/JHEP07(2024)121}{\emph{JHEP} {\bf 07} (2024)
  121}, [\href{http://arxiv.org/abs/2402.00431}{{\tt 2402.00431}}].

\bibitem{Kulesza:2015vda}
A.~Kulesza, L.~Motyka, T.~Stebel and V.~Theeuwes, \emph{{Soft gluon resummation
  for associated $t \bar{t} H$ production at the LHC}},
  \href{http://dx.doi.org/10.1007/JHEP03(2016)065}{\emph{JHEP} {\bf 03} (2016)
  065}, [\href{http://arxiv.org/abs/1509.02780}{{\tt 1509.02780}}].

\bibitem{Broggio:2015lya}
A.~Broggio, A.~Ferroglia, B.~D. Pecjak, A.~Signer and L.~L. Yang,
  \emph{{Associated production of a top pair and a Higgs boson beyond NLO}},
  \href{http://dx.doi.org/10.1007/JHEP03(2016)124}{\emph{JHEP} {\bf 03} (2016)
  124}, [\href{http://arxiv.org/abs/1510.01914}{{\tt 1510.01914}}].

\bibitem{Broggio:2016lfj}
A.~Broggio, A.~Ferroglia, B.~D. Pecjak and L.~L. Yang, \emph{{NNLL resummation
  for the associated production of a top pair and a Higgs boson at the LHC}},
  \href{http://dx.doi.org/10.1007/JHEP02(2017)126}{\emph{JHEP} {\bf 02} (2017)
  126}, [\href{http://arxiv.org/abs/1611.00049}{{\tt 1611.00049}}].

\bibitem{Kulesza:2017ukk}
A.~Kulesza, L.~Motyka, T.~Stebel and V.~Theeuwes, \emph{{Associated $t \bar{t}
  H$ production at the LHC: Theoretical predictions at NLO+NNLL accuracy}},
  \href{http://dx.doi.org/10.1103/PhysRevD.97.114007}{\emph{Phys. Rev. D} {\bf
  97} (2018) 114007}, [\href{http://arxiv.org/abs/1704.03363}{{\tt
  1704.03363}}].

\bibitem{Ju:2019lwp}
W.-L. Ju and L.~L. Yang, \emph{{Resummation of soft and Coulomb corrections for
  $ t\overline{t}h $ production at the LHC}},
  \href{http://dx.doi.org/10.1007/JHEP06(2019)050}{\emph{JHEP} {\bf 06} (2019)
  050}, [\href{http://arxiv.org/abs/1904.08744}{{\tt 1904.08744}}].

\bibitem{Kulesza:2018tqz}
A.~Kulesza, L.~Motyka, D.~Schwartl\"ander, T.~Stebel and V.~Theeuwes,
  \emph{{Associated production of a top quark pair with a heavy electroweak
  gauge boson at NLO$+$NNLL accuracy}},
  \href{http://dx.doi.org/10.1140/epjc/s10052-019-6746-z}{\emph{Eur. Phys. J.
  C} {\bf 79} (2019) 249}, [\href{http://arxiv.org/abs/1812.08622}{{\tt
  1812.08622}}].

\bibitem{Broggio:2019ewu}
A.~Broggio, A.~Ferroglia, R.~Frederix, D.~Pagani, B.~D. Pecjak and I.~Tsinikos,
  \emph{{Top-quark pair hadroproduction in association with a heavy boson at
  NLO+NNLL including EW corrections}},
  \href{http://dx.doi.org/10.1007/JHEP08(2019)039}{\emph{JHEP} {\bf 08} (2019)
  039}, [\href{http://arxiv.org/abs/1907.04343}{{\tt 1907.04343}}].

\bibitem{Kulesza:2020nfh}
A.~Kulesza, L.~Motyka, D.~Schwartl\"ander, T.~Stebel and V.~Theeuwes,
  \emph{{Associated top quark pair production with a heavy boson: differential
  cross sections at NLO+NNLL accuracy}},
  \href{http://dx.doi.org/10.1140/epjc/s10052-020-7987-6}{\emph{Eur. Phys. J.
  C} {\bf 80} (2020) 428}, [\href{http://arxiv.org/abs/2001.03031}{{\tt
  2001.03031}}].

\bibitem{Maltoni:2016yxb}
F.~Maltoni, E.~Vryonidou and C.~Zhang, \emph{{Higgs production in association
  with a top-antitop pair in the Standard Model Effective Field Theory at NLO
  in QCD}}, \href{http://dx.doi.org/10.1007/JHEP10(2016)123}{\emph{JHEP} {\bf
  10} (2016) 123}, [\href{http://arxiv.org/abs/1607.05330}{{\tt 1607.05330}}].

\bibitem{DiNoi:2023onw}
S.~Di~Noi and R.~Gr\"ober, \emph{{Renormalisation group running effects in
  $pp\rightarrow t{\bar{t}}h$ in the Standard Model Effective Field Theory}},
  \href{http://dx.doi.org/10.1140/epjc/s10052-024-12661-5}{\emph{Eur. Phys. J.
  C} {\bf 84} (2024) 403}, [\href{http://arxiv.org/abs/2312.11327}{{\tt
  2312.11327}}].

\bibitem{Frederix:2011zi}
R.~Frederix, S.~Frixione, V.~Hirschi, F.~Maltoni, R.~Pittau and P.~Torrielli,
  \emph{{Scalar and pseudoscalar Higgs production in association with a
  top\textendash{}antitop pair}},
  \href{http://dx.doi.org/10.1016/j.physletb.2011.06.012}{\emph{Phys. Lett. B}
  {\bf 701} (2011) 427--433}, [\href{http://arxiv.org/abs/1104.5613}{{\tt
  1104.5613}}].

\bibitem{Hartanto:2015uka}
H.~B. Hartanto, B.~Jager, L.~Reina and D.~Wackeroth, \emph{{Higgs boson
  production in association with top quarks in the POWHEG BOX}},
  \href{http://dx.doi.org/10.1103/PhysRevD.91.094003}{\emph{Phys. Rev. D} {\bf
  91} (2015) 094003}, [\href{http://arxiv.org/abs/1501.04498}{{\tt
  1501.04498}}].

\bibitem{Denner:2015yca}
A.~Denner and R.~Feger, \emph{{NLO QCD corrections to off-shell top-antitop
  production with leptonic decays in association with a Higgs boson at the
  LHC}}, \href{http://dx.doi.org/10.1007/JHEP11(2015)209}{\emph{JHEP} {\bf 11}
  (2015) 209}, [\href{http://arxiv.org/abs/1506.07448}{{\tt 1506.07448}}].

\bibitem{Stremmer:2021bnk}
D.~Stremmer and M.~Worek, \emph{{Production and decay of the Higgs boson in
  association with top quarks}},
  \href{http://dx.doi.org/10.1007/JHEP02(2022)196}{\emph{JHEP} {\bf 02} (2022)
  196}, [\href{http://arxiv.org/abs/2111.01427}{{\tt 2111.01427}}].

\bibitem{Denner:2016wet}
A.~Denner, J.-N. Lang, M.~Pellen and S.~Uccirati, \emph{{Higgs production in
  association with off-shell top-antitop pairs at NLO EW and QCD at the LHC}},
  \href{http://dx.doi.org/10.1007/JHEP02(2017)053}{\emph{JHEP} {\bf 02} (2017)
  053}, [\href{http://arxiv.org/abs/1612.07138}{{\tt 1612.07138}}].

\bibitem{Demartin:2014fia}
F.~Demartin, F.~Maltoni, K.~Mawatari, B.~Page and M.~Zaro, \emph{{Higgs
  characterisation at NLO in QCD: CP properties of the top-quark Yukawa
  interaction}},
  \href{http://dx.doi.org/10.1140/epjc/s10052-014-3065-2}{\emph{Eur. Phys. J.
  C} {\bf 74} (2014) 3065}, [\href{http://arxiv.org/abs/1407.5089}{{\tt
  1407.5089}}].

\bibitem{Hermann:2022vit}
J.~Hermann, D.~Stremmer and M.~Worek, \emph{{$ \mathcal{CP} $ structure of the
  top-quark Yukawa interaction: NLO QCD corrections and off-shell effects}},
  \href{http://dx.doi.org/10.1007/JHEP09(2022)138}{\emph{JHEP} {\bf 09} (2022)
  138}, [\href{http://arxiv.org/abs/2205.09983}{{\tt 2205.09983}}].

\bibitem{Andersen:2024czj}
J.~Andersen et~al., \emph{{Les Houches 2023: Physics at TeV Colliders: Standard
  Model Working Group Report}},  in \emph{{Physics of the TeV Scale and Beyond
  the Standard Model}: {Intensifying the Quest for New Physics}}, 6, 2024.
\newblock \href{http://arxiv.org/abs/2406.00708}{{\tt 2406.00708}}.

\bibitem{Catani:2007vq}
S.~Catani and M.~Grazzini, \emph{{An NNLO subtraction formalism in hadron
  collisions and its application to Higgs boson production at the LHC}},
  \href{http://dx.doi.org/10.1103/PhysRevLett.98.222002}{\emph{Phys. Rev.
  Lett.} {\bf 98} (2007) 222002},
  [\href{http://arxiv.org/abs/hep-ph/0703012}{{\tt hep-ph/0703012}}].

\bibitem{Catani:2013tia}
S.~Catani, L.~Cieri, D.~de~Florian, G.~Ferrera and M.~Grazzini,
  \emph{{Universality of transverse-momentum resummation and hard factors at
  the NNLO}},
  \href{http://dx.doi.org/10.1016/j.nuclphysb.2014.02.011}{\emph{Nucl. Phys. B}
  {\bf 881} (2014) 414--443}, [\href{http://arxiv.org/abs/1311.1654}{{\tt
  1311.1654}}].

\bibitem{Zhu:2012ts}
H.~X. Zhu, C.~S. Li, H.~T. Li, D.~Y. Shao and L.~L. Yang,
  \emph{{Transverse-momentum resummation for top-quark pairs at hadron
  colliders}},
  \href{http://dx.doi.org/10.1103/PhysRevLett.110.082001}{\emph{Phys. Rev.
  Lett.} {\bf 110} (2013) 082001}, [\href{http://arxiv.org/abs/1208.5774}{{\tt
  1208.5774}}].

\bibitem{Li:2013mia}
H.~T. Li, C.~S. Li, D.~Y. Shao, L.~L. Yang and H.~X. Zhu, \emph{{Top quark pair
  production at small transverse momentum in hadronic collisions}},
  \href{http://dx.doi.org/10.1103/PhysRevD.88.074004}{\emph{Phys. Rev.} {\bf
  D88} (2013) 074004}, [\href{http://arxiv.org/abs/1307.2464}{{\tt
  1307.2464}}].

\bibitem{Catani:2014qha}
S.~Catani, M.~Grazzini and A.~Torre, \emph{{Transverse-momentum resummation for
  heavy-quark hadroproduction}},
  \href{http://dx.doi.org/10.1016/j.nuclphysb.2014.11.019}{\emph{Nucl. Phys.}
  {\bf B890} (2014) 518--538}, [\href{http://arxiv.org/abs/1408.4564}{{\tt
  1408.4564}}].

\bibitem{Bonciani:2015sha}
R.~Bonciani, S.~Catani, M.~Grazzini, H.~Sargsyan and A.~Torre, \emph{{The $q_T$
  subtraction method for top quark production at hadron colliders}},
  \href{http://dx.doi.org/10.1140/epjc/s10052-015-3793-y}{\emph{Eur. Phys. J.}
  {\bf C75} (2015) 581}, [\href{http://arxiv.org/abs/1508.03585}{{\tt
  1508.03585}}].

\bibitem{Catani:2023tby}
S.~Catani, S.~Devoto, M.~Grazzini and J.~Mazzitelli, \emph{{Soft-parton
  contributions to heavy-quark production at low transverse momentum}},
  \href{http://dx.doi.org/10.1007/JHEP04(2023)144}{\emph{JHEP} {\bf 04} (2023)
  144}, [\href{http://arxiv.org/abs/2301.11786}{{\tt 2301.11786}}].

\bibitem{Catani:2019iny}
S.~Catani, S.~Devoto, M.~Grazzini, S.~Kallweit, J.~Mazzitelli and H.~Sargsyan,
  \emph{{Top-quark pair hadroproduction at next-to-next-to-leading order in
  QCD}}, \href{http://dx.doi.org/10.1103/PhysRevD.99.051501}{\emph{Phys. Rev.}
  {\bf D99} (2019) 051501}, [\href{http://arxiv.org/abs/1901.04005}{{\tt
  1901.04005}}].

\bibitem{Catani:2019hip}
S.~Catani, S.~Devoto, M.~Grazzini, S.~Kallweit and J.~Mazzitelli,
  \emph{{Top-quark pair production at the LHC: Fully differential QCD
  predictions at NNLO}},
  \href{http://dx.doi.org/10.1007/JHEP07(2019)100}{\emph{JHEP} {\bf 07} (2019)
  100}, [\href{http://arxiv.org/abs/1906.06535}{{\tt 1906.06535}}].

\bibitem{Catani:2020kkl}
S.~Catani, S.~Devoto, M.~Grazzini, S.~Kallweit and J.~Mazzitelli,
  \emph{{Bottom-quark production at hadron colliders: fully differential
  predictions in NNLO QCD}},
  \href{http://dx.doi.org/10.1007/JHEP03(2021)029}{\emph{JHEP} {\bf 03} (2021)
  029}, [\href{http://arxiv.org/abs/2010.11906}{{\tt 2010.11906}}].

\bibitem{inprepQQXsoft}
S.~Devoto and J.~Mazzitelli, \emph{{in preparation}}, .

\bibitem{Buonocore:2022pqq}
L.~Buonocore, S.~Devoto, S.~Kallweit, J.~Mazzitelli, L.~Rottoli and C.~Savoini,
  \emph{{Associated production of a W boson and massive bottom quarks at
  next-to-next-to-leading order in QCD}},
  \href{http://dx.doi.org/10.1103/PhysRevD.107.074032}{\emph{Phys. Rev. D} {\bf
  107} (2023) 074032}, [\href{http://arxiv.org/abs/2212.04954}{{\tt
  2212.04954}}].

\bibitem{Buonocore:2023ljm}
L.~Buonocore, S.~Devoto, M.~Grazzini, S.~Kallweit, J.~Mazzitelli, L.~Rottoli
  et~al., \emph{{Precise Predictions for the Associated Production of a W Boson
  with a Top-Antitop Quark Pair at the LHC}},
  \href{http://dx.doi.org/10.1103/PhysRevLett.131.231901}{\emph{Phys. Rev.
  Lett.} {\bf 131} (2023) 231901}, [\href{http://arxiv.org/abs/2306.16311}{{\tt
  2306.16311}}].

\bibitem{Ferroglia:2009ii}
A.~Ferroglia, M.~Neubert, B.~D. Pecjak and L.~L. Yang, \emph{{Two-loop
  divergences of massive scattering amplitudes in non-abelian gauge theories}},
  \href{http://dx.doi.org/10.1088/1126-6708/2009/11/062}{\emph{JHEP} {\bf 11}
  (2009) 062}, [\href{http://arxiv.org/abs/0908.3676}{{\tt 0908.3676}}].

\bibitem{Barnreuther:2013qvf}
P.~B\"arnreuther, M.~Czakon and P.~Fiedler, \emph{{Virtual amplitudes and
  threshold behaviour of hadronic top-quark pair-production cross sections}},
  \href{http://dx.doi.org/10.1007/JHEP02(2014)078}{\emph{JHEP} {\bf 02} (2014)
  078}, [\href{http://arxiv.org/abs/1312.6279}{{\tt 1312.6279}}].

\bibitem{Penin:2005eh}
A.~A. Penin, \emph{{Two-loop photonic corrections to massive Bhabha
  scattering}},
  \href{http://dx.doi.org/10.1016/j.nuclphysb.2005.11.016}{\emph{Nucl. Phys. B}
  {\bf 734} (2006) 185--202}, [\href{http://arxiv.org/abs/hep-ph/0508127}{{\tt
  hep-ph/0508127}}].

\bibitem{Mitov:2006xs}
A.~Mitov and S.~Moch, \emph{{The Singular behavior of massive QCD amplitudes}},
  \href{http://dx.doi.org/10.1088/1126-6708/2007/05/001}{\emph{JHEP} {\bf 05}
  (2007) 001}, [\href{http://arxiv.org/abs/hep-ph/0612149}{{\tt
  hep-ph/0612149}}].

\bibitem{Becher:2007cu}
T.~Becher and K.~Melnikov, \emph{{Two-loop QED corrections to Bhabha
  scattering}},
  \href{http://dx.doi.org/10.1088/1126-6708/2007/06/084}{\emph{JHEP} {\bf 06}
  (2007) 084}, [\href{http://arxiv.org/abs/0704.3582}{{\tt 0704.3582}}].

\bibitem{Engel:2018fsb}
T.~Engel, C.~Gnendiger, A.~Signer and Y.~Ulrich, \emph{{Small-mass effects in
  heavy-to-light form factors}},
  \href{http://dx.doi.org/10.1007/JHEP02(2019)118}{\emph{JHEP} {\bf 02} (2019)
  118}, [\href{http://arxiv.org/abs/1811.06461}{{\tt 1811.06461}}].

\bibitem{Wang:2023qbf}
G.~Wang, T.~Xia, L.~L. Yang and X.~Ye, \emph{{On the high-energy behavior of
  massive QCD amplitudes}},
  \href{http://dx.doi.org/10.1007/JHEP05(2024)082}{\emph{JHEP} {\bf 05} (2024)
  082}, [\href{http://arxiv.org/abs/2312.12242}{{\tt 2312.12242}}].

\bibitem{Badger:2021ega}
S.~Badger, H.~B. Hartanto, J.~Kry\'s and S.~Zoia, \emph{{Two-loop
  leading-colour QCD helicity amplitudes for Higgs boson production in
  association with a bottom-quark pair at the LHC}},
  \href{http://dx.doi.org/10.1007/JHEP11(2021)012}{\emph{JHEP} {\bf 11} (2021)
  012}, [\href{http://arxiv.org/abs/2107.14733}{{\tt 2107.14733}}].

\bibitem{Badger:2024awe}
S.~Badger, H.~B. Hartanto, R.~Poncelet, Z.~Wu, Y.~Zhang and S.~Zoia,
  \emph{{Full-colour double-virtual amplitudes for associated production of a
  Higgs boson with a bottom-quark pair at the LHC}},
  \href{http://arxiv.org/abs/2412.06519}{{\tt 2412.06519}}.

\bibitem{Grazzini:2017mhc}
M.~Grazzini, S.~Kallweit and M.~Wiesemann, \emph{{Fully differential NNLO
  computations with MATRIX}},
  \href{http://dx.doi.org/10.1140/epjc/s10052-018-5771-7}{\emph{Eur. Phys. J.
  C} {\bf 78} (2018) 537}, [\href{http://arxiv.org/abs/1711.06631}{{\tt
  1711.06631}}].

\bibitem{Catani:1996jh}
S.~Catani and M.~H. Seymour, \emph{{The Dipole formalism for the calculation of
  QCD jet cross-sections at next-to-leading order}},
  \href{http://dx.doi.org/10.1016/0370-2693(96)00425-X}{\emph{Phys. Lett. B}
  {\bf 378} (1996) 287--301}, [\href{http://arxiv.org/abs/hep-ph/9602277}{{\tt
  hep-ph/9602277}}].

\bibitem{Catani:1996vz}
S.~Catani and M.~H. Seymour, \emph{{A General algorithm for calculating jet
  cross-sections in NLO QCD}},
  \href{http://dx.doi.org/10.1016/S0550-3213(96)00589-5}{\emph{Nucl. Phys. B}
  {\bf 485} (1997) 291--419}, [\href{http://arxiv.org/abs/hep-ph/9605323}{{\tt
  hep-ph/9605323}}].

\bibitem{Catani:2002hc}
S.~Catani, S.~Dittmaier, M.~H. Seymour and Z.~Trocsanyi, \emph{{The Dipole
  formalism for next-to-leading order QCD calculations with massive partons}},
  \href{http://dx.doi.org/10.1016/S0550-3213(02)00098-6}{\emph{Nucl. Phys. B}
  {\bf 627} (2002) 189--265}, [\href{http://arxiv.org/abs/hep-ph/0201036}{{\tt
  hep-ph/0201036}}].

\bibitem{Kallweit:2017khh}
S.~Kallweit, J.~M. Lindert, S.~Pozzorini and M.~Sch\"onherr, \emph{{NLO QCD+EW
  predictions for $2\ell2\nu$ diboson signatures at the LHC}},
  \href{http://dx.doi.org/10.1007/JHEP11(2017)120}{\emph{JHEP} {\bf 11} (2017)
  120}, [\href{http://arxiv.org/abs/1705.00598}{{\tt 1705.00598}}].

\bibitem{Dittmaier:1999mb}
S.~Dittmaier, \emph{{A General approach to photon radiation off fermions}},
  \href{http://dx.doi.org/10.1016/S0550-3213(99)00563-5}{\emph{Nucl. Phys. B}
  {\bf 565} (2000) 69--122}, [\href{http://arxiv.org/abs/hep-ph/9904440}{{\tt
  hep-ph/9904440}}].

\bibitem{Dittmaier:2008md}
S.~Dittmaier, A.~Kabelschacht and T.~Kasprzik, \emph{{Polarized QED splittings
  of massive fermions and dipole subtraction for non-collinear-safe
  observables}},
  \href{http://dx.doi.org/10.1016/j.nuclphysb.2008.03.010}{\emph{Nucl. Phys. B}
  {\bf 800} (2008) 146--189}, [\href{http://arxiv.org/abs/0802.1405}{{\tt
  0802.1405}}].

\bibitem{Gehrmann:2010ry}
T.~Gehrmann and N.~Greiner, \emph{{Photon Radiation with MadDipole}},
  \href{http://dx.doi.org/10.1007/JHEP12(2010)050}{\emph{JHEP} {\bf 12} (2010)
  050}, [\href{http://arxiv.org/abs/1011.0321}{{\tt 1011.0321}}].

\bibitem{Schonherr:2017qcj}
M.~Sch\"onherr, \emph{{An automated subtraction of NLO EW infrared
  divergences}},
  \href{http://dx.doi.org/10.1140/epjc/s10052-018-5600-z}{\emph{Eur. Phys. J.
  C} {\bf 78} (2018) 119}, [\href{http://arxiv.org/abs/1712.07975}{{\tt
  1712.07975}}].

\bibitem{Cascioli:2011va}
F.~Cascioli, P.~Maierhofer and S.~Pozzorini, \emph{{Scattering Amplitudes with
  Open Loops}},
  \href{http://dx.doi.org/10.1103/PhysRevLett.108.111601}{\emph{Phys. Rev.
  Lett.} {\bf 108} (2012) 111601}, [\href{http://arxiv.org/abs/1111.5206}{{\tt
  1111.5206}}].

\bibitem{Buccioni:2017yxi}
F.~Buccioni, S.~Pozzorini and M.~Zoller, \emph{{On-the-fly reduction of open
  loops}}, \href{http://dx.doi.org/10.1140/epjc/s10052-018-5562-1}{\emph{Eur.
  Phys. J. C} {\bf 78} (2018) 70}, [\href{http://arxiv.org/abs/1710.11452}{{\tt
  1710.11452}}].

\bibitem{Buccioni:2019sur}
F.~Buccioni, J.-N. Lang, J.~M. Lindert, P.~Maierh\"ofer, S.~Pozzorini, H.~Zhang
  et~al., \emph{{OpenLoops 2}},
  \href{http://dx.doi.org/10.1140/epjc/s10052-019-7306-2}{\emph{Eur. Phys. J.
  C} {\bf 79} (2019) 866}, [\href{http://arxiv.org/abs/1907.13071}{{\tt
  1907.13071}}].

\bibitem{Actis:2012qn}
S.~Actis, A.~Denner, L.~Hofer, A.~Scharf and S.~Uccirati, \emph{{Recursive
  generation of one-loop amplitudes in the Standard Model}},
  \href{http://dx.doi.org/10.1007/JHEP04(2013)037}{\emph{JHEP} {\bf 04} (2013)
  037}, [\href{http://arxiv.org/abs/1211.6316}{{\tt 1211.6316}}].

\bibitem{Actis:2016mpe}
S.~Actis, A.~Denner, L.~Hofer, J.-N. Lang, A.~Scharf and S.~Uccirati,
  \emph{{RECOLA: REcursive Computation of One-Loop Amplitudes}},
  \href{http://dx.doi.org/10.1016/j.cpc.2017.01.004}{\emph{Comput. Phys.
  Commun.} {\bf 214} (2017) 140--173},
  [\href{http://arxiv.org/abs/1605.01090}{{\tt 1605.01090}}].

\bibitem{Denner:2017wsf}
A.~Denner, J.-N. Lang and S.~Uccirati, \emph{{Recola2: REcursive Computation of
  One-Loop Amplitudes 2}},
  \href{http://dx.doi.org/10.1016/j.cpc.2017.11.013}{\emph{Comput. Phys.
  Commun.} {\bf 224} (2018) 346--361},
  [\href{http://arxiv.org/abs/1711.07388}{{\tt 1711.07388}}].

\bibitem{Becher:2007ty}
T.~Becher, M.~Neubert and G.~Xu, \emph{{Dynamical Threshold Enhancement and
  Resummation in Drell-Yan Production}},
  \href{http://dx.doi.org/10.1088/1126-6708/2008/07/030}{\emph{JHEP} {\bf 07}
  (2008) 030}, [\href{http://arxiv.org/abs/0710.0680}{{\tt 0710.0680}}].

\bibitem{Becher:2014oda}
T.~Becher, A.~Broggio and A.~Ferroglia, \emph{{Introduction to Soft-Collinear
  Effective Theory}}, vol.~896.
\newblock Springer, 2015,
  \href{http://dx.doi.org/10.1007/978-3-319-14848-9}{10.1007/978-3-319-14848-9}.

\bibitem{Ahrens:2010zv}
V.~Ahrens, A.~Ferroglia, M.~Neubert, B.~D. Pecjak and L.~L. Yang,
  \emph{{Renormalization-Group Improved Predictions for Top-Quark Pair
  Production at Hadron Colliders}},
  \href{http://dx.doi.org/10.1007/JHEP09(2010)097}{\emph{JHEP} {\bf 09} (2010)
  097}, [\href{http://arxiv.org/abs/1003.5827}{{\tt 1003.5827}}].

\bibitem{Broggio:2016zgg}
A.~Broggio, A.~Ferroglia, G.~Ossola and B.~D. Pecjak, \emph{{Associated
  production of a top pair and a W boson at next-to-next-to-leading logarithmic
  accuracy}}, \href{http://dx.doi.org/10.1007/JHEP09(2016)089}{\emph{JHEP} {\bf
  09} (2016) 089}, [\href{http://arxiv.org/abs/1607.05303}{{\tt 1607.05303}}].

\bibitem{Broggio:2017kzi}
A.~Broggio, A.~Ferroglia, G.~Ossola, B.~D. Pecjak and R.~D. Sameshima,
  \emph{{Associated production of a top pair and a Z boson at the LHC to NNLL
  accuracy}}, \href{http://dx.doi.org/10.1007/JHEP04(2017)105}{\emph{JHEP} {\bf
  04} (2017) 105}, [\href{http://arxiv.org/abs/1702.00800}{{\tt 1702.00800}}].

\bibitem{Czakon:2018nun}
M.~Czakon, A.~Ferroglia, D.~Heymes, A.~Mitov, B.~D. Pecjak, D.~J. Scott et~al.,
  \emph{{Resummation for (boosted) top-quark pair production at NNLO+NNLL' in
  QCD}}, \href{http://dx.doi.org/10.1007/JHEP05(2018)149}{\emph{JHEP} {\bf 05}
  (2018) 149}, [\href{http://arxiv.org/abs/1803.07623}{{\tt 1803.07623}}].

\bibitem{Catani:1996yz}
S.~Catani, M.~L. Mangano, P.~Nason and L.~Trentadue, \emph{{The Resummation of
  soft gluons in hadronic collisions}},
  \href{http://dx.doi.org/10.1016/0550-3213(96)00399-9}{\emph{Nucl. Phys. B}
  {\bf 478} (1996) 273--310}, [\href{http://arxiv.org/abs/hep-ph/9604351}{{\tt
  hep-ph/9604351}}].

\bibitem{Catani:1989ne}
S.~Catani and L.~Trentadue, \emph{{Resummation of the QCD Perturbative Series
  for Hard Processes}},
  \href{http://dx.doi.org/10.1016/0550-3213(89)90273-3}{\emph{Nucl. Phys. B}
  {\bf 327} (1989) 323--352}.

\bibitem{Sterman:1986aj}
G.~F. Sterman, \emph{{Summation of Large Corrections to Short Distance Hadronic
  Cross-Sections}},
  \href{http://dx.doi.org/10.1016/0550-3213(87)90258-6}{\emph{Nucl. Phys. B}
  {\bf 281} (1987) 310--364}.

\bibitem{Bonciani:1998vc}
R.~Bonciani, S.~Catani, M.~L. Mangano and P.~Nason, \emph{{NLL resummation of
  the heavy quark hadroproduction cross-section}},
  \href{http://dx.doi.org/10.1016/S0550-3213(98)00335-6}{\emph{Nucl. Phys. B}
  {\bf 529} (1998) 424--450}, [\href{http://arxiv.org/abs/hep-ph/9801375}{{\tt
  hep-ph/9801375}}].

\bibitem{Kidonakis:1997gm}
N.~Kidonakis and G.~F. Sterman, \emph{{Resummation for QCD hard scattering}},
  \href{http://dx.doi.org/10.1016/S0550-3213(97)00506-3}{\emph{Nucl. Phys. B}
  {\bf 505} (1997) 321--348}, [\href{http://arxiv.org/abs/hep-ph/9705234}{{\tt
  hep-ph/9705234}}].

\bibitem{Catani:2003zt}
S.~Catani, D.~de~Florian, M.~Grazzini and P.~Nason, \emph{{Soft gluon
  resummation for Higgs boson production at hadron colliders}},
  \href{http://dx.doi.org/10.1088/1126-6708/2003/07/028}{\emph{JHEP} {\bf 07}
  (2003) 028}, [\href{http://arxiv.org/abs/hep-ph/0306211}{{\tt
  hep-ph/0306211}}].

\bibitem{Ferroglia:2009ep}
A.~Ferroglia, M.~Neubert, B.~D. Pecjak and L.~L. Yang, \emph{{Two-loop
  divergences of scattering amplitudes with massive partons}},
  \href{http://dx.doi.org/10.1103/PhysRevLett.103.201601}{\emph{Phys. Rev.
  Lett.} {\bf 103} (2009) 201601}, [\href{http://arxiv.org/abs/0907.4791}{{\tt
  0907.4791}}].

\bibitem{Alwall:2014hca}
J.~Alwall, R.~Frederix, S.~Frixione, V.~Hirschi, F.~Maltoni, O.~Mattelaer
  et~al., \emph{{The automated computation of tree-level and next-to-leading
  order differential cross sections, and their matching to parton shower
  simulations}}, \href{http://dx.doi.org/10.1007/JHEP07(2014)079}{\emph{JHEP}
  {\bf 07} (2014) 079}, [\href{http://arxiv.org/abs/1405.0301}{{\tt
  1405.0301}}].

\bibitem{Czakon:2017wor}
M.~Czakon, D.~Heymes, A.~Mitov, D.~Pagani, I.~Tsinikos and M.~Zaro,
  \emph{{Top-pair production at the LHC through NNLO QCD and NLO EW}},
  \href{http://dx.doi.org/10.1007/JHEP10(2017)186}{\emph{JHEP} {\bf 10} (2017)
  186}, [\href{http://arxiv.org/abs/1705.04105}{{\tt 1705.04105}}].

\bibitem{Pagani:2016caq}
D.~Pagani, I.~Tsinikos and M.~Zaro, \emph{{The impact of the photon PDF and
  electroweak corrections on $t \bar{t}$ distributions}},
  \href{http://dx.doi.org/10.1140/epjc/s10052-016-4318-z}{\emph{Eur. Phys. J.
  C} {\bf 76} (2016) 479}, [\href{http://arxiv.org/abs/1606.01915}{{\tt
  1606.01915}}].

\bibitem{PDF4LHCWorkingGroup:2022cjn}
{\scshape PDF4LHC Working Group} collaboration, R.~D. Ball et~al., \emph{{The
  PDF4LHC21 combination of global PDF fits for the LHC Run III}},
  \href{http://dx.doi.org/10.1088/1361-6471/ac7216}{\emph{J. Phys. G} {\bf 49}
  (2022) 080501}, [\href{http://arxiv.org/abs/2203.05506}{{\tt 2203.05506}}].

\bibitem{Manohar:2016nzj}
A.~Manohar, P.~Nason, G.~P. Salam and G.~Zanderighi, \emph{{How bright is the
  proton? A precise determination of the photon parton distribution function}},
  \href{http://dx.doi.org/10.1103/PhysRevLett.117.242002}{\emph{Phys. Rev.
  Lett.} {\bf 117} (2016) 242002}, [\href{http://arxiv.org/abs/1607.04266}{{\tt
  1607.04266}}].

\bibitem{Manohar:2017eqh}
A.~V. Manohar, P.~Nason, G.~P. Salam and G.~Zanderighi, \emph{{The Photon
  Content of the Proton}},
  \href{http://dx.doi.org/10.1007/JHEP12(2017)046}{\emph{JHEP} {\bf 12} (2017)
  046}, [\href{http://arxiv.org/abs/1708.01256}{{\tt 1708.01256}}].

\bibitem{Butterworth:2015oua}
J.~Butterworth et~al., \emph{{PDF4LHC recommendations for LHC Run II}},
  \href{http://dx.doi.org/10.1088/0954-3899/43/2/023001}{\emph{J. Phys. G} {\bf
  43} (2016) 023001}, [\href{http://arxiv.org/abs/1510.03865}{{\tt
  1510.03865}}].

\bibitem{Carrazza:2020gss}
S.~Carrazza, E.~R. Nocera, C.~Schwan and M.~Zaro, \emph{{PineAPPL: combining EW
  and QCD corrections for fast evaluation of LHC processes}},
  \href{http://dx.doi.org/10.1007/JHEP12(2020)108}{\emph{JHEP} {\bf 12} (2020)
  108}, [\href{http://arxiv.org/abs/2008.12789}{{\tt 2008.12789}}].

\bibitem{Dittmaier:2003ej}
S.~Dittmaier, M.~Kr\"amer and M.~Spira, \emph{{Higgs radiation off bottom
  quarks at the Tevatron and the CERN LHC}},
  \href{http://dx.doi.org/10.1103/PhysRevD.70.074010}{\emph{Phys. Rev. D} {\bf
  70} (2004) 074010}, [\href{http://arxiv.org/abs/hep-ph/0309204}{{\tt
  hep-ph/0309204}}].

\bibitem{Wiesemann:2014ioa}
M.~Wiesemann, R.~Frederix, S.~Frixione, V.~Hirschi, F.~Maltoni and
  P.~Torrielli, \emph{{Higgs production in association with bottom quarks}},
  \href{http://dx.doi.org/10.1007/JHEP02(2015)132}{\emph{JHEP} {\bf 02} (2015)
  132}, [\href{http://arxiv.org/abs/1409.5301}{{\tt 1409.5301}}].

\bibitem{Pagani:2020rsg}
D.~Pagani, H.-S. Shao and M.~Zaro, \emph{{RIP $ Hb\overline{b} $: how other
  Higgs production modes conspire to kill a rare signal at the LHC}},
  \href{http://dx.doi.org/10.1007/JHEP11(2020)036}{\emph{JHEP} {\bf 11} (2020)
  036}, [\href{http://arxiv.org/abs/2005.10277}{{\tt 2005.10277}}].

\bibitem{Saibel:2021krs}
A.~Saibel, S.-O. Moch and M.~Aldaya~Martin, \emph{{Cross-sections for ttH
  production with the top quark MSbar mass}},
  \href{http://dx.doi.org/10.1016/j.physletb.2022.137195}{\emph{Phys. Lett. B}
  {\bf 832} (2022) 137195}, [\href{http://arxiv.org/abs/2111.12505}{{\tt
  2111.12505}}].

\end{thebibliography}\endgroup

\end{document}